\documentclass{aa}  

\usepackage{graphicx}
\usepackage{txfonts}
\usepackage{lipsum}
\usepackage{subcaption}         
\usepackage{lscape}             
\usepackage{placeins}           
\usepackage[x11names]{xcolor}
\usepackage{booktabs}
\usepackage{hyperref}

\begin{document}

    \title{Streak detection in the VST/OmegaCAM archive using deep learning}

    \author{E. Rachith\inst{1}
        \and S. Hellmich\inst{1}
        \and V. Fiszbin\inst{1}
        \and B.Y. Irureta-Goyena\inst{1}
        \and A. Price\inst{1}
        \and J.-P. Kneib\inst{1}
        }

   \institute{Laboratory of Astrophysics, EPFL, Chemin Pegasi 51, 1290 Versoix, Switzerland\\
    \email{elisabeth.rachith@epfl.ch}}

   \date{Received January 13, 2026. Accepted June 15, 2026.}

   \authorrunning{E. Rachith et al.}
 
  \abstract  
   {Ground-based astronomical surveys inadvertently capture streaks from satellites and space debris crossing their fields of view. These incidental observations from wide-field instruments such as OmegaCAM on the VST offer valuable opportunities to characterise resident space objects without the need for dedicated observing time.}
   {We developed an automated deep-learning pipeline to detect and classify streaks in the OmegaCAM archive, enabling large-scale analyses of space object populations and their impact on astronomical data.}
   {The pipeline combines an adapted Hough transform lookup-based convolutional neural network (HT-LCNN) for initial streak detection on raw images with a VGG6-based CNN classifier to reject false positives. We augmented a manually annotated dataset of 384\,000 patches from archive images with physically simulated streaks. Following a detection, we applied astrometric calibration and cross-matched the results with the space-track catalogue.}
   {We find the detector achieves F1-scores of 0.966 (validation) and 0.958 (test) on the augmented dataset, detecting $>95\%$ of artificial streaks with a signal-to-noise ratio of $\text{S/N} >4$. On real 2023 data, the precision drops to 0.783 due to image variability, but the classifier boosts it to 0.990, while retaining 97\% of true positives and rejecting $>96\%$ of false positives. Applied to one year of VST observations (1\,246\,048 OmegaCAM CCD frames), the pipeline identified 25\,335 streaks, including more than 20\% uncorrelated with catalogue entries; finally, 16.9\% of images revealed some level of contamination.}
   {The pipeline demonstrates robust performance on real archival data and successfully uncovers faint uncatalogued objects, highlighting the potential of survey archives for debris monitoring.}

   \keywords{streak detection -- 
                space debris characterisation --
                machine learning --
                ground-based telescope
               }

   \maketitle
\nolinenumbers

\section{Introduction}\label{sec:introduction}
In the context of an increasingly congested orbital environment, continuous monitoring of the space debris population is essential. Although most objects larger than a few centimetres are routinely tracked, their physical characteristics and long-term evolution often remain poorly known \citep{horstmann_flux_2021, space_debris_office_esas_2025}. However, understanding these properties is critical for improving debris evolution models \citep{braun_exploiting_2019}, predicting uncontrolled re-entry events \citep{piergentili_leo_2020}, and planning active debris removal missions \citep{okamoto_observational_2025}.

Although commercial space situational awareness (SSA) providers routinely detect and track debris, their services primarily target operational satellites and high‑value objects; debris observations are rarely catalogued or made accessible for scientific use. In parallel, dedicated surveys have greatly advanced our understanding of the space debris population, yet they remain constrained by operational costs, limited observation windows, and a narrow tracking capacity, often focusing on a single object or a small number of targets at a time \citep{mokhnatkin_performance_2017, silha_space_2020, schildknecht_recent_2024}. In contrast, ground-based \citep{bassa_analytical_2022, barentine_aggregate_2023} and even space-based \citep{kruk_impact_2023, goncalves_space-based_2025} astronomical survey telescopes routinely and inadvertently capture reflected light from satellites and space debris as they pass through their fields of view. Wide field instruments such as VST/OmegaCAM \citep{kuijken_omegacam_2011}, Blanco/DECam \citep{flaugher_dark_2015}, ZTF \citep{bellm_zwicky_2019}, Pan-STARRS \citep{tonry_pan-starrs1_2012}, and others observe dozens of such crossings each night. The trails left by these objects on the high quality and high resolution images provide valuable information about their properties. By exploiting these incidental observations, we can derive useful insights without requiring dedicated observation time and since most survey databases are openly accessible, at no additional acquisition cost.

An essential first step in analysing the observed objects is the detection of streaks in the images, which is the main focus of this paper. Several approaches have been developed for this purpose, often primarily designed to remove streaks that interfere with astronomical observations. One such approach is matched filtering \citep{turin_introduction_1960}, which involves convolving the image with a streak template, typically a Gaussian point spread function (PSF) extended along a line, and identifying peaks in the convolved response that reveal the presence of streaks \citep{vananti_improved_2020, cvrcek_fast_2021}. Although this method is capable of detecting very faint streaks, it is computationally demanding and requires a careful optimisation of the filters.

Computer vision techniques such as the Hough \citep{duda_use_1972} and Radon \citep{radon_determination_1986} transforms have also been widely employed to detect satellite traces \citep{wijnen_using_2019, danarianto_prototype_2022}. These methods are robust at identifying linear features, but are prone to false positives arising from unrelated linear structures within images. Despite recent efforts to reduce their computational cost \citep{nir_optimal_2018}, these techniques still require long processing times and, thus, they are not ideal for large scale surveys comprising thousands of images.

In recent years, machine-learning methods have been successfully introduced and consistently outperform traditional approaches in robustness and inference speed, particularly when applied to large datasets \citep{guo_dim_2022, jeffries_detection_2023, liu_identification_2025}. Some studies have proposed hybrid techniques that combine modern deep-learning models with conventional algorithms, for instance, by integrating U-nets with the Hough transform \citep{stoppa_automated_2024} or line segment detectors \citep{chen_artificial_2025}, achieving promising results on specific datasets. Related methods have also been used to identify shorter streaks, such as those produced by near-Earth objects, in wide field astronomical images \citep{duev_deepstreaks_2019, urechiatu_ensemble_2023, irureta-goyena_method_2025}. However, false positives remain a common limitation that continues to affect the overall reliability of such techniques.

Intuitive and easy-to-implement thresholds can remove some false positives. The orientation, length, and intensity profile of a streak, for example, can help reject saturation bleeds, some cosmic rays, and other known image artefacts. However, a more thorough characterisation of false positives and the underlying patterns is required to exclude them as effectively and reliably as possible. \cite{cvrcek_fast_2021} developed a Bayesian framework that filters false detections using thresholds derived from their matched-filtering streak detection method. More recently, machine-learning approaches have gained increasing attention. \cite{duev_deepstreaks_2019} introduced an ensemble of convolutional neural network (CNN) classifiers to robustly discard false positives in asteroid searches with ZTF, while \cite{ntagiou_space_2023} proposed a lightweight multilayer perceptron classifier to sort false detections in the search of space debris streaks.

Building on recent developments, we present a detection routine that combines deep learning with the Hough transform through the Hough transform lookup‑based convolutional neural network (HT-LCNN) method (\citealt{lin_deep_2020}), complemented by a CNN classifier that effectively eliminates false positives. Our objective is to extract physical information about space debris, such as rotation rates and their temporal evolution. To achieve this, we process our detections so that they can later be used for precise photometric analysis and to derive lightcurves.

The remainder of this paper is organised as follows. Section~\ref{sec:data} provides a brief overview of the data used in this study, as well as the selection and preparation of the datasets. Section~\ref{sec:methods} describes the algorithms developed for the detection and classification of streaks. Section~\ref{sec:results} presents the results of the analysis, including the training and validation of the algorithms and their application to one year of VST/OmegaCAM data. Finally, Sect.~\ref{sec:discussion} discusses the results and their implications. Our algorithms and datasets are available on GitHub\footnote{\url{https://github.com/elisabethrachith/odle}} and Zenodo\footnote{\url{https://zenodo.org/records/18097841}}.

\section{Data}\label{sec:data}
\begin{figure}
     \centering
     \includegraphics[width=\linewidth]{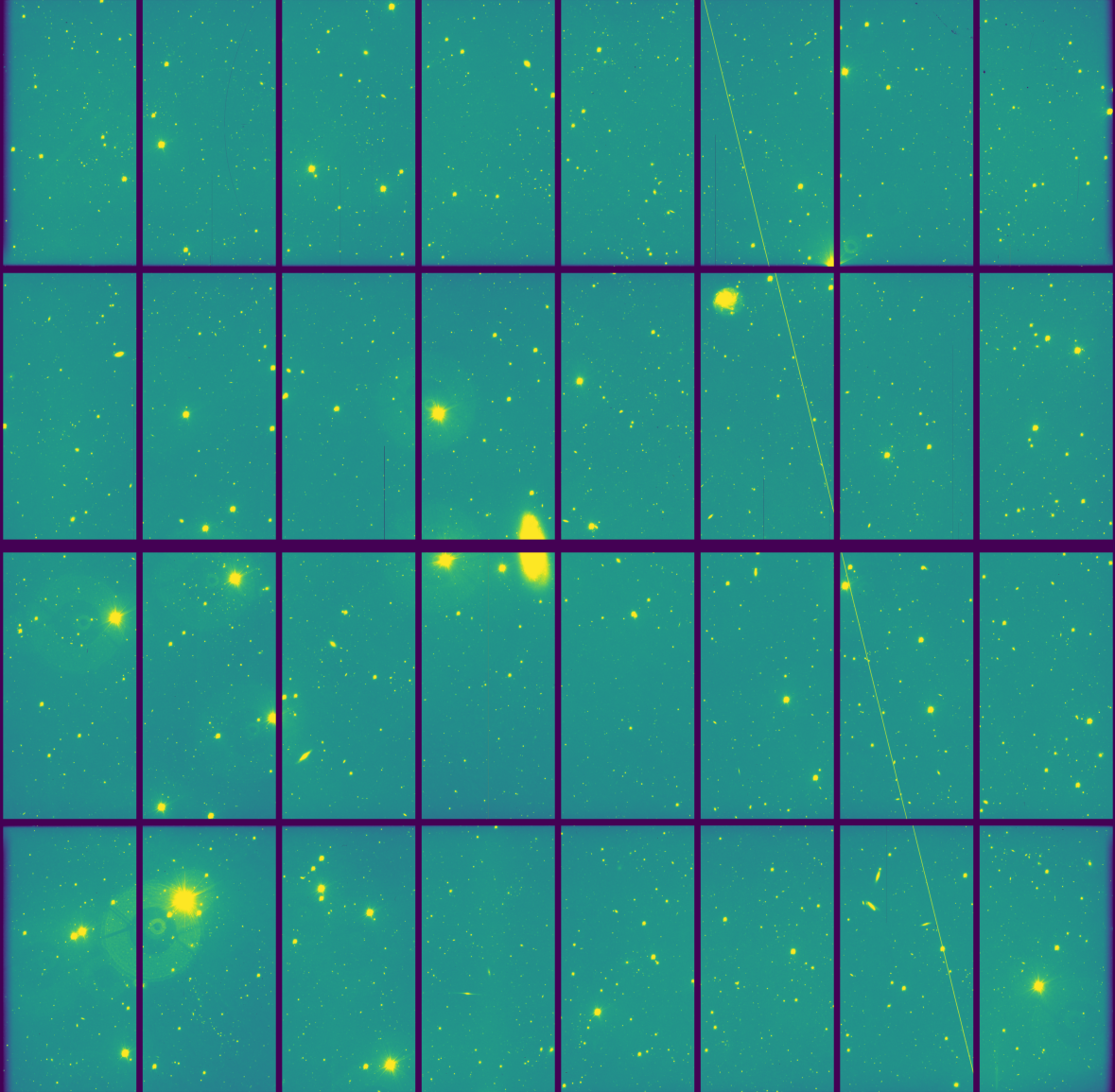}
     \caption{OmegaCAM detector layout with a visible satellite streak.}
     \label{fig:omegacam_exposure}
 \end{figure}

Although our techniques can be applied to different telescope archives, we decided to set our initial focus on the archival data of the VST/OmegaCAM. OmegaCAM is the imager mounted on the 2.6 m VST telescope located at Cerro Paranal in Chile. The instrument’s data are openly accessible\footnote{\url{https://archive.eso.org}}, with an archive spanning more than a decade since 2011. This provides excellent opportunities to analyse the long-term evolution of satellites and space debris. The telescope’s high sensitivity, regularly reaching a limiting magnitude of $m_{AB}>22$ in various configurations \citep{kuijken_gravitational_2015, shanks_vlt_2015}, allows for the detection of very faint streaks that might be missed by commonly used wide-field SSA telescopes. Additionally, with a 16\,000\,$\times$\,16\,000 pixel sensor and a fine pixel scale of 0.21 arcseconds per pixel, OmegaCAM makes it possible to carry out a precise lightcurve reconstruction, which is an essential step towards characterising the physical attributes of detected objects. The instrument captures one square degree of sky per exposure, allowing it to record multiple streaks in a single image. Exposure times vary from a few seconds to 35 minutes with a median of 90\,s. OmegaCAM is equipped with 12 different filters, including a Sloan ugriz set, Johnson B and V filters, several narrow-band filters, and a Strömgren v filter. Figure~\ref{fig:omegacam_exposure} shows an example r-band exposure of OmegaCAM.

\subsection{Streak detection dataset}
For the training, validation and testing of our detection algorithm, we randomly selected 375 images from the entire OmegaCAM archive. Due to the one-year proprietary period at the time of selection, images from 2023 and later were excluded. To ensure sufficient diversity, images from all available filters were included. No general photometric reduction pipeline exists for the images acquired with OmegaCAM, and consequently, no freely accessible dataset of reduced images exists either. As our goal is to analyse the entire OmegaCAM archive containing more than 400\,000 individual exposures, we aimed to make our detection pipeline as efficient as possible. For this reason, we chose to apply the detection algorithm on raw images and only calibrate images with detected streaks at a later stage, before extracting the lightcurves. Given the large size of OmegaCAM mosaics (32 CCDs per image) and their high pixel count, subcrops measuring 512\,$\times$\,512 pixels were prepared from the individual CCD frames and saved as 16-bit monochrome PNG files. This size was found to provide a good compromise between keeping subcrops large enough to provide sufficient contextual information to the network while limiting computational cost, as larger inputs require significantly more processing time. We chose to create subcrops rather than resize the images to preserve the original resolution and avoid any loss of fine detail that resizing might introduce. Instead of using 16-bit FITS files directly, the 16-bit PNG input format was adopted to better integrate with the machine learning library, PyTorch. Pytorch is optimised for standard image formats like PNG, whereas FITS files resulted in slower data loading. Of the 375 selected images, 305 were allocated for training the algorithm, corresponding to a total of 312\,320 individual PNG subcrops. The remaining images were divided equally between the validation and test sets, with 35 images each, resulting in 35\,840 subcrops per set. The streaks in the selected images were manually annotated by defining start and endpoints using Label Studio \citep{tkachenko_label_2025}. Given the limited number of real streaks found in the dataset (1\,861 streaks across 384\,000 subcrops) and the high time cost of manual annotation, we generated artificial streaks to augment the dataset and better evaluate the algorithm's performance.

\subsection{Artificial streak generation}
To ensure that the algorithm is able to generalise effectively across the OmegaCAM archive, it is essential that the artificially generated streaks closely resemble real streaks observed in the images. We began by generating an artificial population of objects selected randomly from a predefined pool of sizes and orbital altitudes. As very large objects tend to have a high signal-to-noise ratio (S/N) and are easily detected by the algorithm, we focused on smaller objects with sizes between 1 cm and 2 m on circular orbits. The altitudes ranged from 300 km to 36\,100 km. The pool included a higher proportion of smaller objects located near the Earth's surface, reflecting the characteristics of the actual space object population, in particular space debris. The complete set of available sizes and altitudes used in the selection is given in Appendix~\ref{sec:app_snr_derivation}.

A corresponding streak was generated for each of the selected objects by rendering lines with varying lengths and orientations to simulate object motion across the detector. Most OmegaCAM images have exposure times that are long enough for streaks to extend across the entire field of view. To reflect this behaviour, 80\% of the simulated streaks span the full width or height of the detector, while 20\% have one endpoint located within the detector boundaries.

To compute the local S/N of each streak, we first estimated the sky background brightness. We applied the sigma-clipping algorithm from the Astropy library \citep{the_astropy_collaboration_astropy_2022} to determine an appropriate threshold for the identification of sources. We applied circular masks to the detected sources to exclude them from the background calculation. Finally, we applied sigma-clipping statistics to the remaining unmasked pixels and used the mean value as the sky background brightness. We computed the flux from each object by assuming a spherical shape with a diameter taken from the predefined population and using Lambertian light scattering. We derived the theoretical apparent magnitude using the Sun’s magnitude as a reference, accounting for the object’s phase function, albedo, atmospheric extinction at the observed wavelength, and observer-target range. From the apparent magnitude, we estimated the total flux contribution of the object's reflected light using the flux of a magnitude 20 source as a reference, including the effects of local seeing, exposure time, streak length, and imager characteristics. This calculation provided the local S/N of the object. The mathematical derivation is given in Appendix~\ref{sec:app_snr_derivation}. We discarded streaks with $\text{S/N}< 2$ as they lacked sufficient information for lightcurve analysis or meaningful training data.

\begin{figure}[h!]
    \centering
    \includegraphics[width=\linewidth]{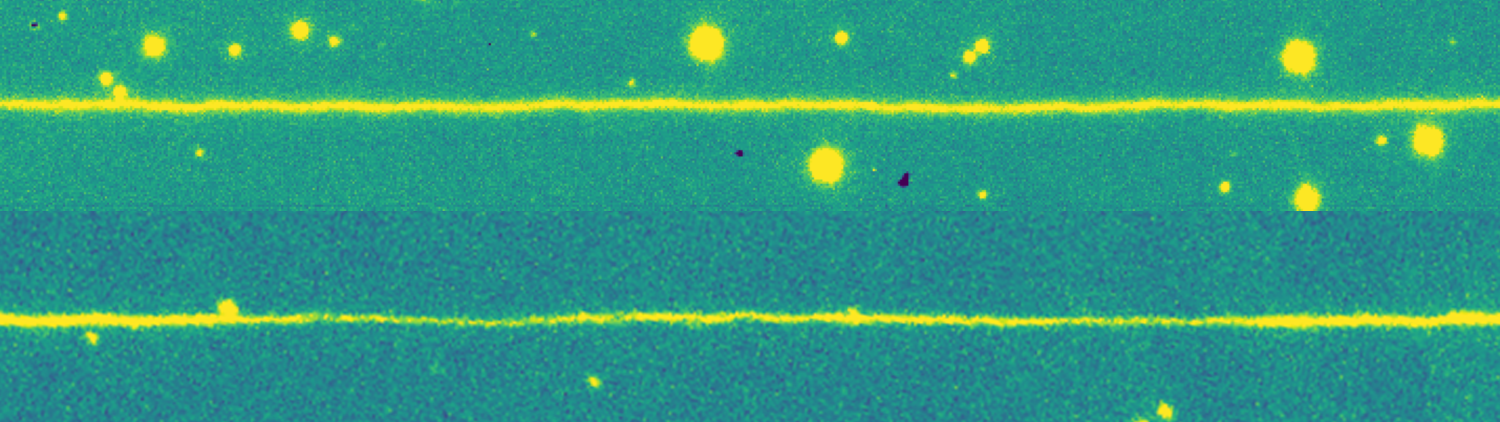}
    \caption{Simulated streak (top) and real streak (bottom) of objects in GEO.}
    \label{fig:wobbling_comparison}
\end{figure}

Accurate simulations of satellite streaks require the  modelling of both the object motion and atmospheric effects. The motion of the satellite or space debris produces the linear shape of the trace, while atmospheric effects such as scintillation and seeing alter its brightness, degrade its sharpness, and introduce deviations from the linear shape. In long exposures, such as those in the OmegaCAM archive, seeing usually dominates for fixed objects and causes blurring. For fast-moving objects crossing the field of view, scintillation becomes locally significant and introduces rapid intensity fluctuations along the streak. Although the rotation of objects can produce dashed or varying-intensity streaks, we did not model this effect in the present study. We modelled atmospheric scintillation effects using Langevin dynamics, a stochastic framework that describes systems influenced by both deterministic forces and random noise, allowing scintillation to be seen as a random walk inside a potential well. This approach effectively captures the random fluctuations caused by atmospheric turbulence while maintaining a stable equilibrium around a central position. Because seeing effects can be described by a Moffat profile, we used this profile as the harmonic potential in our Langevin dynamics framework. A detailed explanation and the mathematical derivation of the scintillation simulation is provided in Appendix~\ref{sec:app_scintillation}. The impact of atmospheric scintillation on a simulated and a real streak is shown in Fig.~\ref{fig:wobbling_comparison}. We generated a total of 224\,291 artificial streaks, which were inserted throughout the 384\,000 subcrops of our dataset.

\subsection{Classifier dataset}
After experimenting with several input formats, we found that 8-bit greyscale PNG images with Z-scale normalisation gave the best classifier performance. We generate these images using the output of the detection algorithm, which provides the endpoint locations of the streaks. First, we rotate the original images to horizontally align the streaks using the method described in Sect.~\ref{sec:detection_refinement}. We then create crops around the streaks with a fixed height of 200 pixels and a length that varies according to the detected streak length. After generating the crops, we resize them to 144\,$\times$\,144 pixels as this input format yielded the best results in our ablation studies. An example image used as input by the classifier is shown in Fig.~\ref{fig:classifier_data}. Further details and supporting plots are available in Appendix~\ref{sec:app_classifier}.

\begin{figure}[h!]
    \centering
    \begin{subfigure}[b]{\linewidth}
        \centering
        \includegraphics[width=\linewidth]{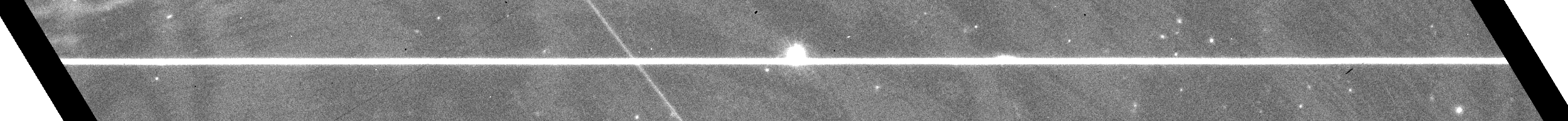}
        \vspace{0.2pt}
    \end{subfigure}
    \begin{subfigure}[b]{\linewidth}
        \centering
        \includegraphics[width=0.3\linewidth]{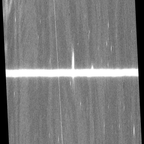}
    \end{subfigure}
    \caption{Sample input data used by the classification algorithm. Streak cutout before (top) and after (bottom) resizing.}
    \label{fig:classifier_data}
\end{figure}

\begin{figure}[h!]
    \centering
    \includegraphics[width=\linewidth]{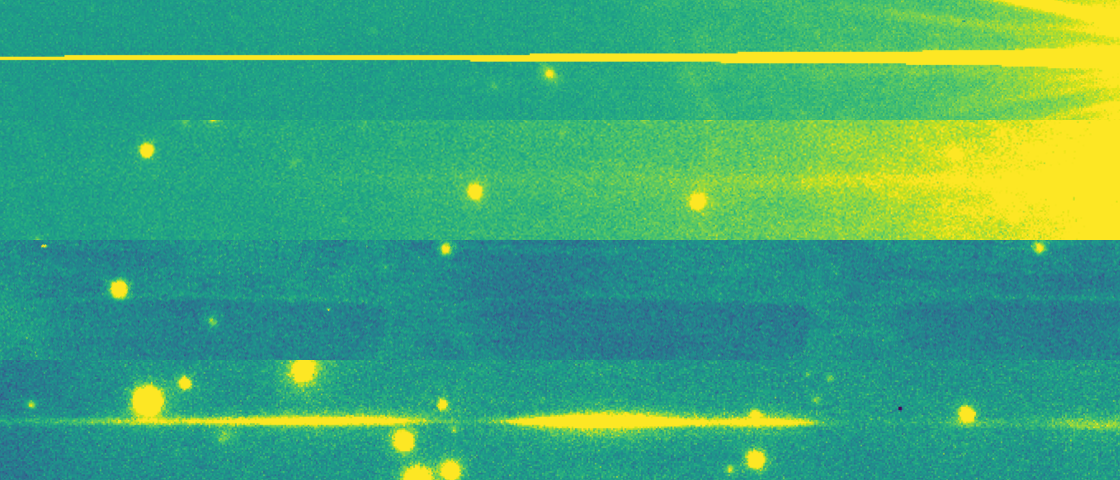}
    \caption{Examples of common false-positive detections made by the detection algorithm. Bleeding (top), diffraction spikes (centre top), filter fringes (centre bottom), and miscellaneous (bottom).}
    \label{fig:fp_samples}
\end{figure}

We selected the training, validation, and test datasets from data collected in 2023, providing an unseen evaluation setting for the detection algorithm which was trained on pre-2023 data. We applied the detection method to the months of January to March and November and December 2023, and manually reviewed the detections to classify them as true or false positives. We further divided the false positives into four common categories: bleeding from saturated stars, diffraction spikes from bright stars, filter fringes, and a miscellaneous category for detections that did not fit the previous groups. Examples of each false-positive category are presented in Fig.~\ref{fig:fp_samples}. We used the January -- March data for training the classifier, with a total of 10\,783 streaks, and validated the model on the November data, which contained 4\,199 streaks. The December data, with 3\,370 streaks, was reserved for testing the method.

\section{Methods}\label{sec:methods}
This section presents the methodology developed for the detection of streaks in OmegaCAM images, the subsequent refinement of detections, astrometric calibration and the cross-correlation with an object catalogue. We also present the architecture of the classification algorithm used to identify and reject false positives.

\subsection{Streak detection with HT-LCNN}
We adapted the HT‑LCNN algorithm by \cite{lin_deep_2020} for astronomical imaging to perform the initial streak detection. Integrating the Hough transform within the network improves the recognition of geometric features, such as satellite streaks and their characteristic linear shape. The Hough transform is incorporated into the network via a residual branch. Featuremaps from the initial convolutional layers are transformed into the Hough domain, where they are processed with local convolutions that capture linear structures across the image. The results are then projected back into image space and merged with the standard convolutional branch, allowing the network to integrate both local image information and global geometric structure. This Hough-transform residual block can be integrated into any CNN and is not limited to the global implementation used here. An architecture schematic is provided in Appendix~\ref{sec:app_htlcnn_schema}. We modified the model to handle 16‑bit monochrome images, standard in astronomy, to benefit from the greater dynamic range compared with the 8‑bit images typically used in conventional machine-learning tasks.

Using small 512\,$\times$\,512 pixel crops allows us to reliably detect streaks while maintaining reasonable inference times, but this scale is impractical for later steps such as orbit fitting and lightcurve extraction. We reconstruct the streaks within each CCD tile by first removing duplicates and discarding detections shorter than five pixels. We then map the remaining detections back to their positions on the CCD and compute their Hough parameters $\rho$ (distance from the origin to the nearest point of the line) and $\theta$ (angle between the X‑axis and the line) to describe their geometry. We group lines with similar parameters, using thresholds of 50 pixels for $\rho$ and 0.1 radians for $\theta$; these values were chosen to maximise the CCD-level F1-score on the validation set. We then select the most distant endpoints in each cluster to define the stitched streak segment. As the use of small images may produce more false positives by detecting features that look like streaks in isolation but not in the larger context, the stitching step also enables us to filter out part of these false detections by discarding singleton segments that do not cluster with any others. Although this could occasionally exclude short streaks located at the very corner of an image, we accept this trade-off: genuine features will typically appear in adjacent CCDs, allowing us to study them from neighbouring data, while those that cannot be recovered are typically too short to contribute meaningfully to subsequent analyses. No stitching across CCDs is performed in the current study. Because gaps exist between CCDs and only single raw images are used for streak detection, these gaps cannot be reconstructed, and subsequent analyses such as photometric extraction can therefore only be performed at the CCD level. We also discard images with poor telescope tracking which results in trailing stars that are misidentified as streaks by the network. We identify poor tracking as when all three of the following conditions are met: (i) more than 30 streaks are detected across all CCDs, (ii) more than 25 CCDs contain streaks, and (iii) the mean number of streaks per CCD exceeds four.

\subsection{Detection refinement}\label{sec:detection_refinement}
The start and endpoints returned by the network do not always precisely align with the streak in the image. Thus, we refined them before going on to obtain accurate astrometry measurements.

\begin{figure*}[h]
    \centering
    \includegraphics[width=\linewidth]{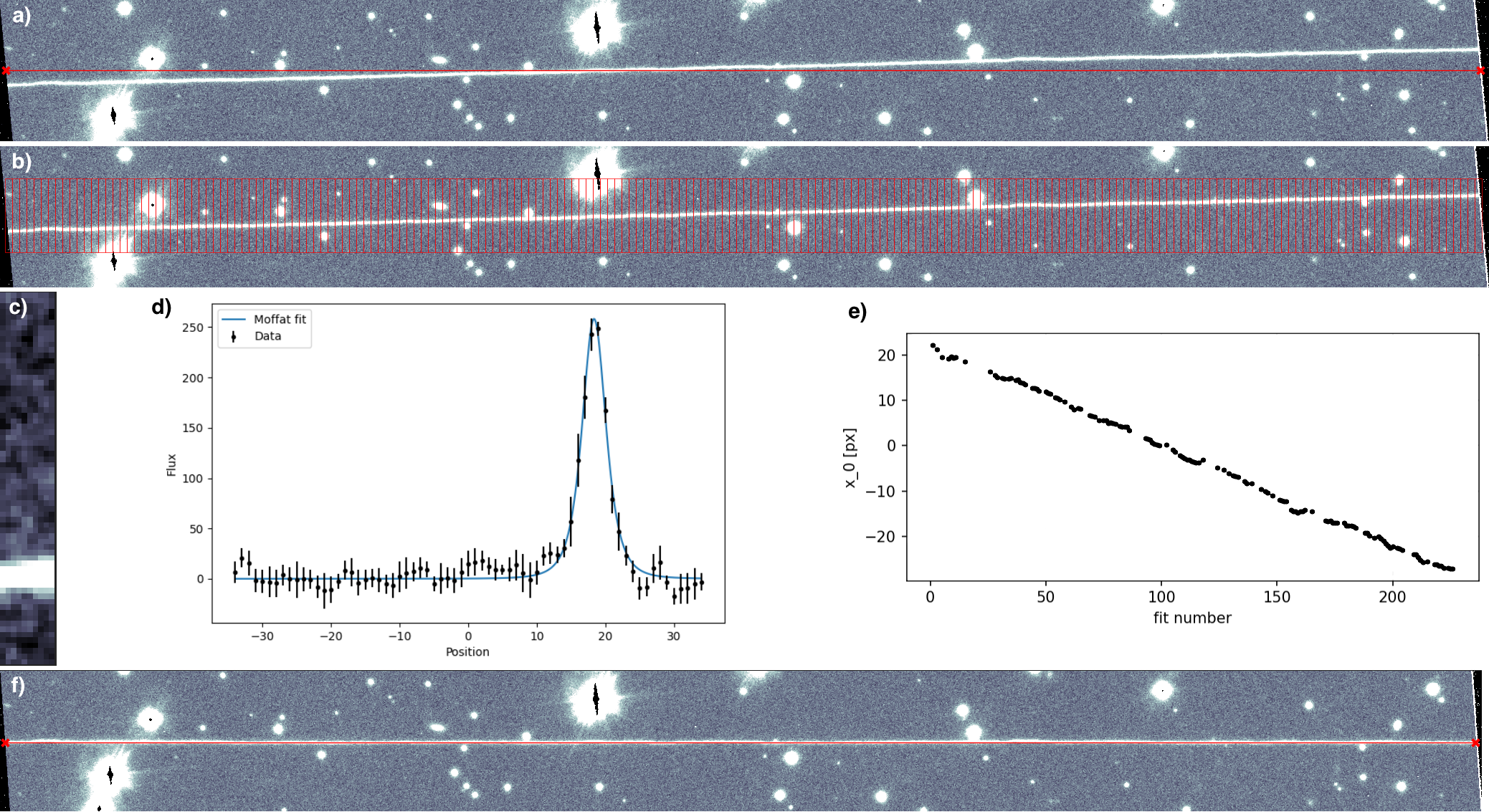}
    \caption{Streak alignment procedure. (a) Original image cutout showing a detection from the network. (b) Rotated cutout with the streak approximately horizontal and divided into equal‑length segments. (c) Zoomed view of the streak showing the extraction region used for the profiles. (d) Example cross‑sectional profile of a segment with the Moffat model fit. (e) Fitted streak centre positions along the track as a function of segment index. (f) Final aligned streak with updated endpoints projected onto the fitted line.}
    \label{fig:streak_alignment}
\end{figure*}

For each detection, we cut out a small image and rotate it so that the streak is roughly horizontal. We then split the streak into equal‑length segments and, for each segment, median‑combine the pixels along the horizontal direction to obtain a cross‑sectional profile. Each profile is fitted with a Moffat function, and the peak of the fit gives the centre of the streak in that segment. A straight line is then fitted through all centres, and the original endpoints are projected onto this line to better match the actual streak. This procedure is repeated until the fitted slope is smaller than 0.1 divided by the streak length in pixels. The procedure is illustrated in Fig.~\ref{fig:streak_alignment}.

The segment width along the streak is set to 1.4 times the full width at half maximum (FWHM) of the Gaussian PSF measured from guide stars on the auxiliary OmegaCAM CCDs. The segment height is first chosen to allow for a maximum angle difference of three degrees between the detected and true streak, and is then reduced in each alignment iteration to avoid including background stars that could disturb the Moffat fit. After the alignment is complete, the refined endpoints are transformed back into the original image coordinates. For faint streaks with low S/N, the segments are widened until the Moffat fit converges or the width reaches 20 divided by the streak length in pixels. If no fit is found within this limit, the detection is marked as too faint for reliable processing or as a possible false positive.

\subsection{Astrometric calibration and object correlation}
The raw OmegaCAM images include an initial astrometric solution that must be refined for high‑precision astrometric applications. We improve the astrometric solution using Astrometry.net \citep{lang_astrometrynet_2010} to plate solve the images. We then use this refined calibration to derive accurate right ascension and declination for the start and endpoints of the detected streaks.

Next, we matched the detected streaks with known satellites and space debris listed in the publicly available space‑track\footnote{\url{https://www.space-track.org/}} catalogue. For each observation, we propagate the two‑line elements (TLEs) closest in time to the image timestamp to predict which objects may have crossed the detector during acquisition. We then project these candidates onto the image and compare them with the detected streaks. Uncertainties from the TLEs and propagation errors can result in the projection not exactly matching the observation. Therefore, we select the best candidate based on the minimum angular separation and position‑angle difference between the predicted and observed streaks.

\subsection{Classifier}\label{sec:methods-classifier}
\begin{figure}[h!]
    \centering
    \includegraphics[width=\linewidth]{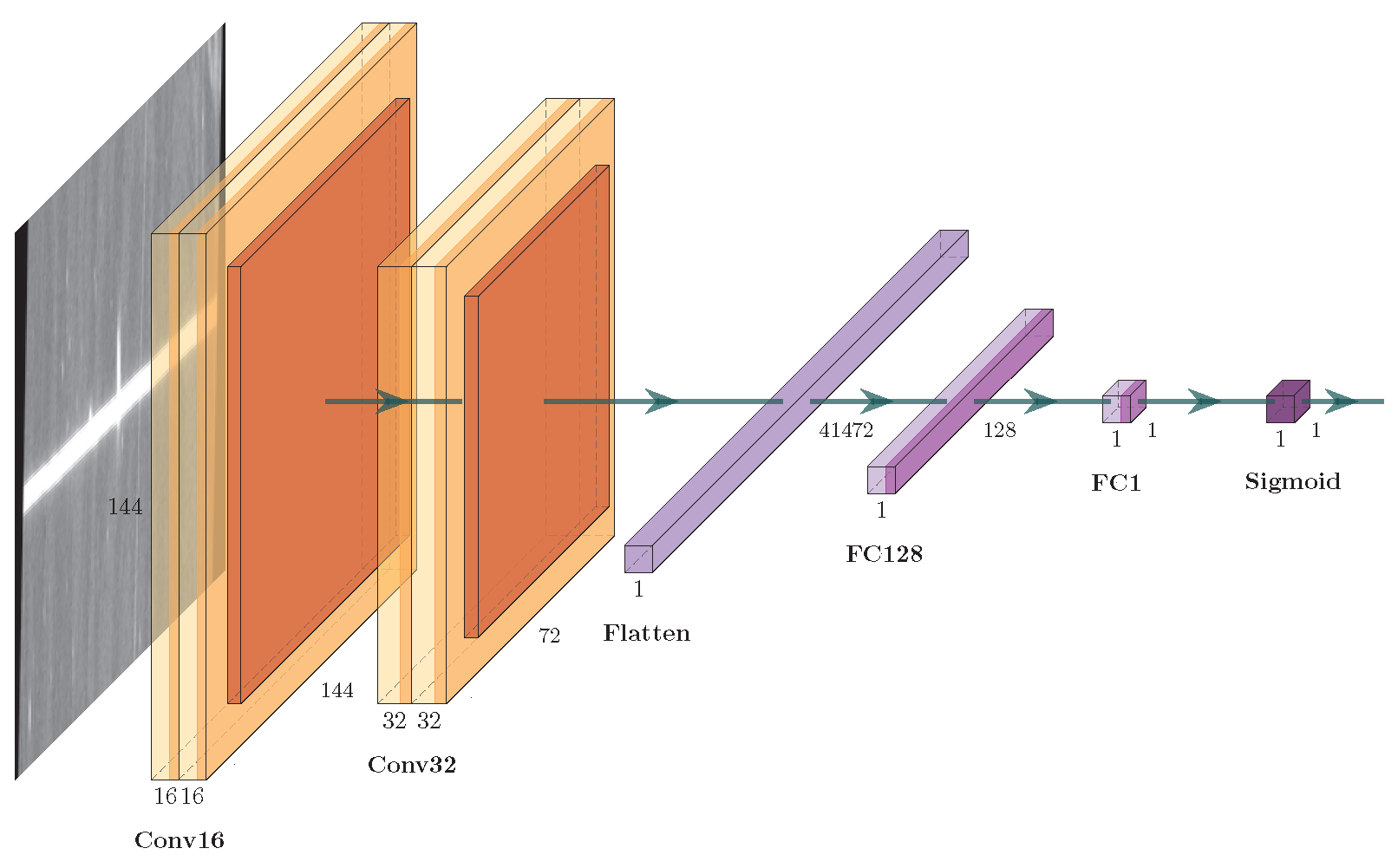}
    \caption{Schematic representation of the CNN used for binary streak classification. The network takes an 8-bit single-channel 144\,$\times$\,144 pixel input image and passes it through two convolutional blocks (yellow) with batch normalisation and ReLU activations (orange), each followed by max-pooling (red) and 0.25 dropout for regularisation. The extracted features are flattened and passed to the classifier, which consists of two fully connected layers (purple) with sigmoid activation (dark purple) and 0.5 dropout between them. The final magenta box represents the output probability of the ‘true streak’ class.}
    \label{fig:vgg6}
\end{figure}

The classifier architecture is adapted from the VGG6 deep-learning network as implemented in DeepStreaks by \cite{duev_deepstreaks_2019}. It consists of a feature extractor followed by a binary classifier that outputs the probability that a given streak is a true positive detection. The architecture of the network is illustrated in Fig.~\ref{fig:vgg6}.

As a baseline, we also evaluate a coordinate-only model using the refined streak endpoints provided by the pipeline, in both image (pixel) and sky (right ascension/declination) coordinates. The multi-layer perceptron (MLP) that we trained on this endpoint-only data is a fully connected neural network that takes tabular inputs based on eight dimensions. It consists of three dense hidden layers with dimensions 64, 32, and 16, with ReLU activations used after each layer and a dropout layer with rate 0.1 applied after the first hidden layer. The final linear layer outputs a single scalar, which is passed through a Sigmoid activation to produce a probability score for binary classification.

\section{Results}\label{sec:results}
This section presents the results obtained from the training and validation of the detection algorithm, demonstrating its performance and generalisation to real-world OmegaCAM data. It describes the training and validation of the classification algorithm and concludes with the application of the complete methodology to one year of archival observations.

\subsection{Training and performance of the detection algorithm}
\begin{figure}[h!]
    \centering
    \begin{subfigure}[b]{\linewidth}
        \centering
        \includegraphics[width=\linewidth]{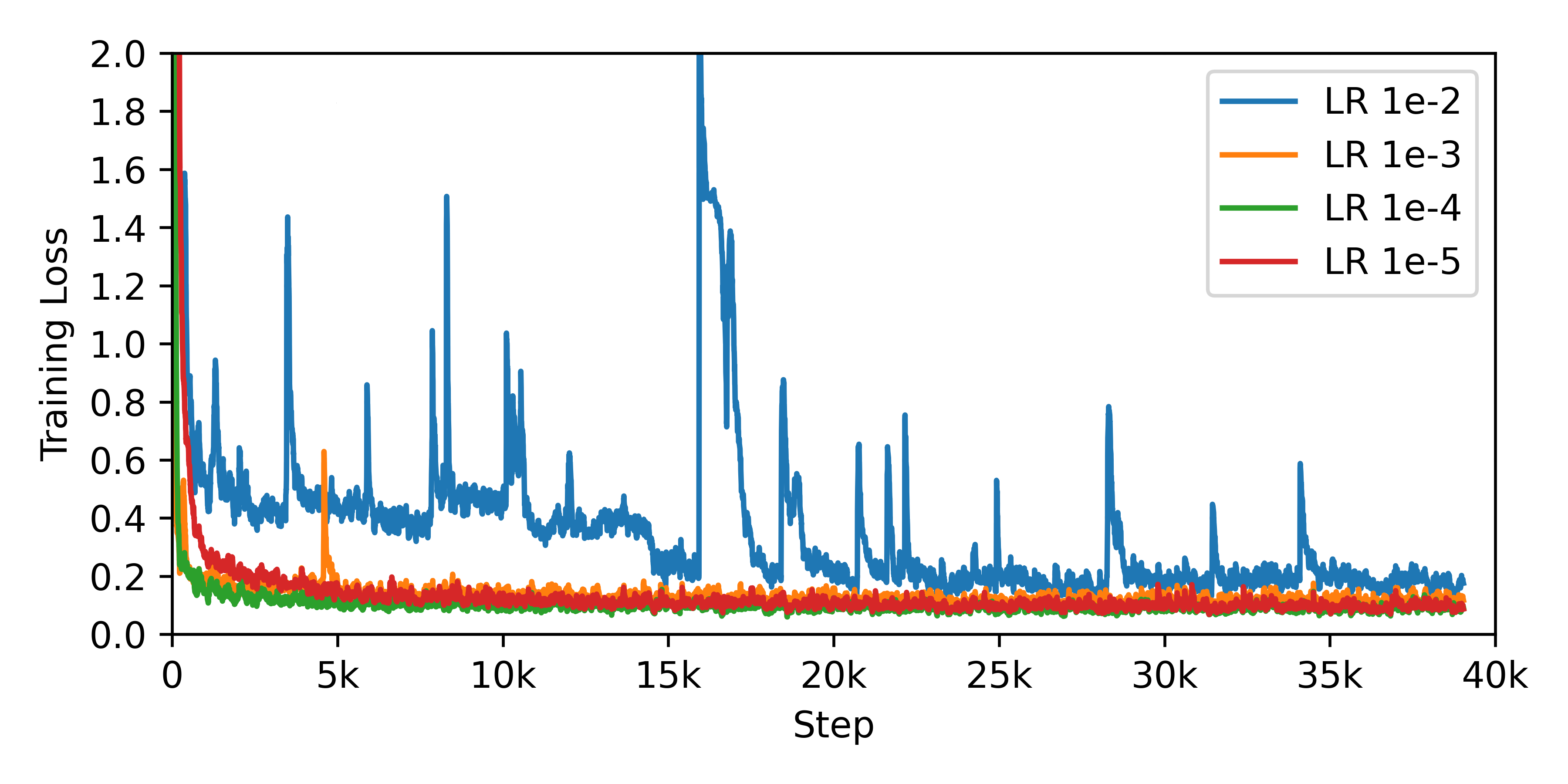}
        \caption{}
        \label{fig:training_loss_analysis}
    \end{subfigure}
    \begin{subfigure}[b]{\linewidth}
        \centering
        \includegraphics[width=\linewidth]{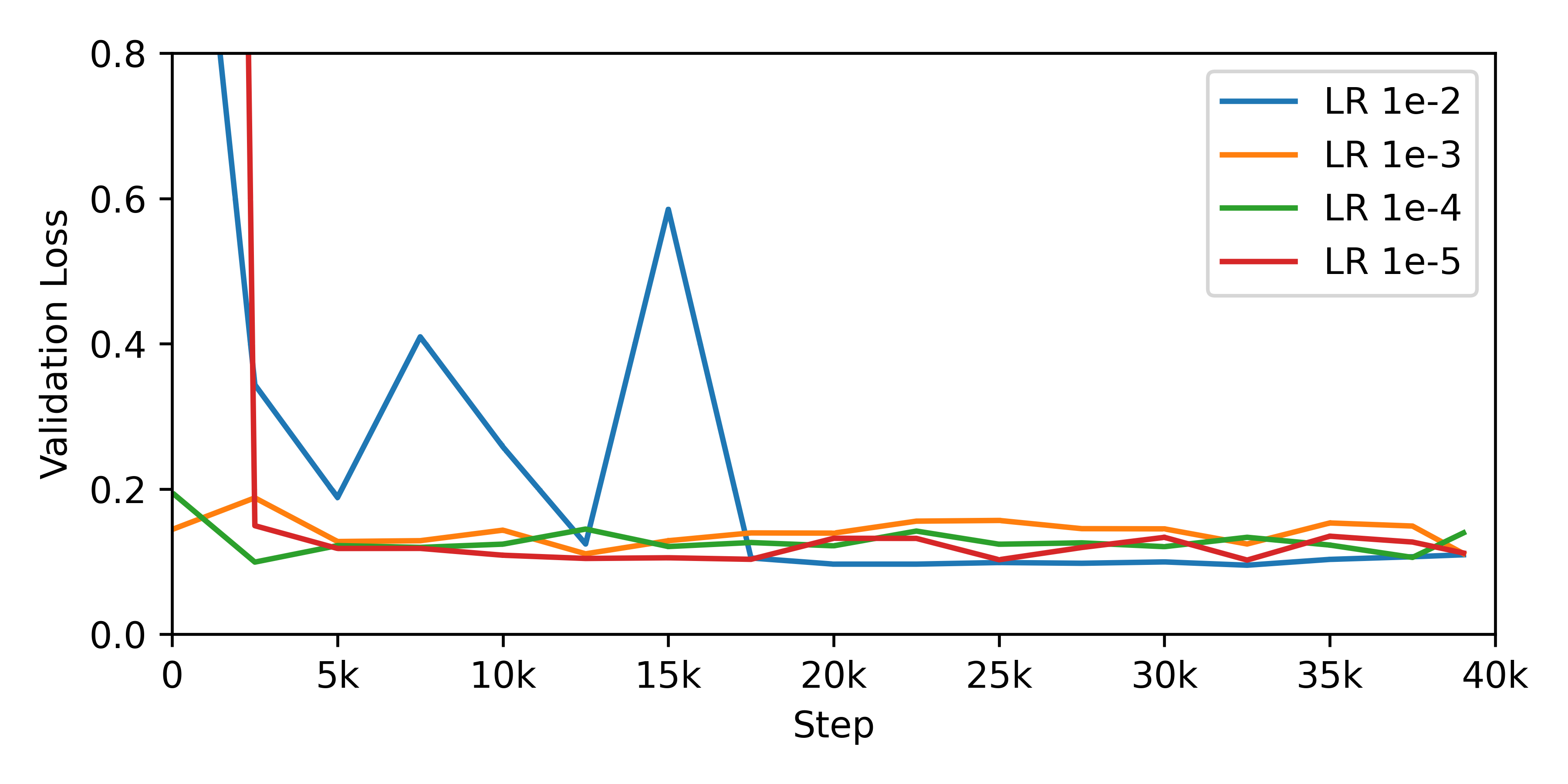}
        \caption{}
        \label{fig:validation_loss_analysis}
    \end{subfigure}
    \caption{Detection algorithm training loss (a) and validation loss (b) curves for different learning rates (one epoch each). }
    \label{fig:lr_analysis}
\end{figure}

Before training the final algorithm, we performed an initial learning‑rate search to find the optimal value for our configuration and dataset. We trained the model for one epoch at various learning rates and analysed the training and validation losses, as can be seen in Fig.~\ref{fig:lr_analysis}. One epoch was sufficient due to the large size of the dataset, but memory limits restricted the batch size to eight samples. To reduce the noise that this small batch size introduced into the loss curves, we smoothed the training loss using an exponential moving average.

The results showed that a learning rate of $1\times 10^{-2}$ was too high: the training loss lacked a steep initial descent, displayed instability with visible peaks, and remained high, while the validation loss was also highly unstable. At the other extreme, $1\times 10^{-5}$ achieved a better training loss but still showed an insufficiently steep initial descent. Intermediate values of $1\times 10^{-3}$ and $1\times 10^{-4}$ produced better results, with $1\times 10^{-4}$ performing best by delivering lower training loss, slightly better validation loss, and greater stability than $1\times 10^{-3}$.

\begin{figure}[h!]
    \centering
    \includegraphics[width=\linewidth]{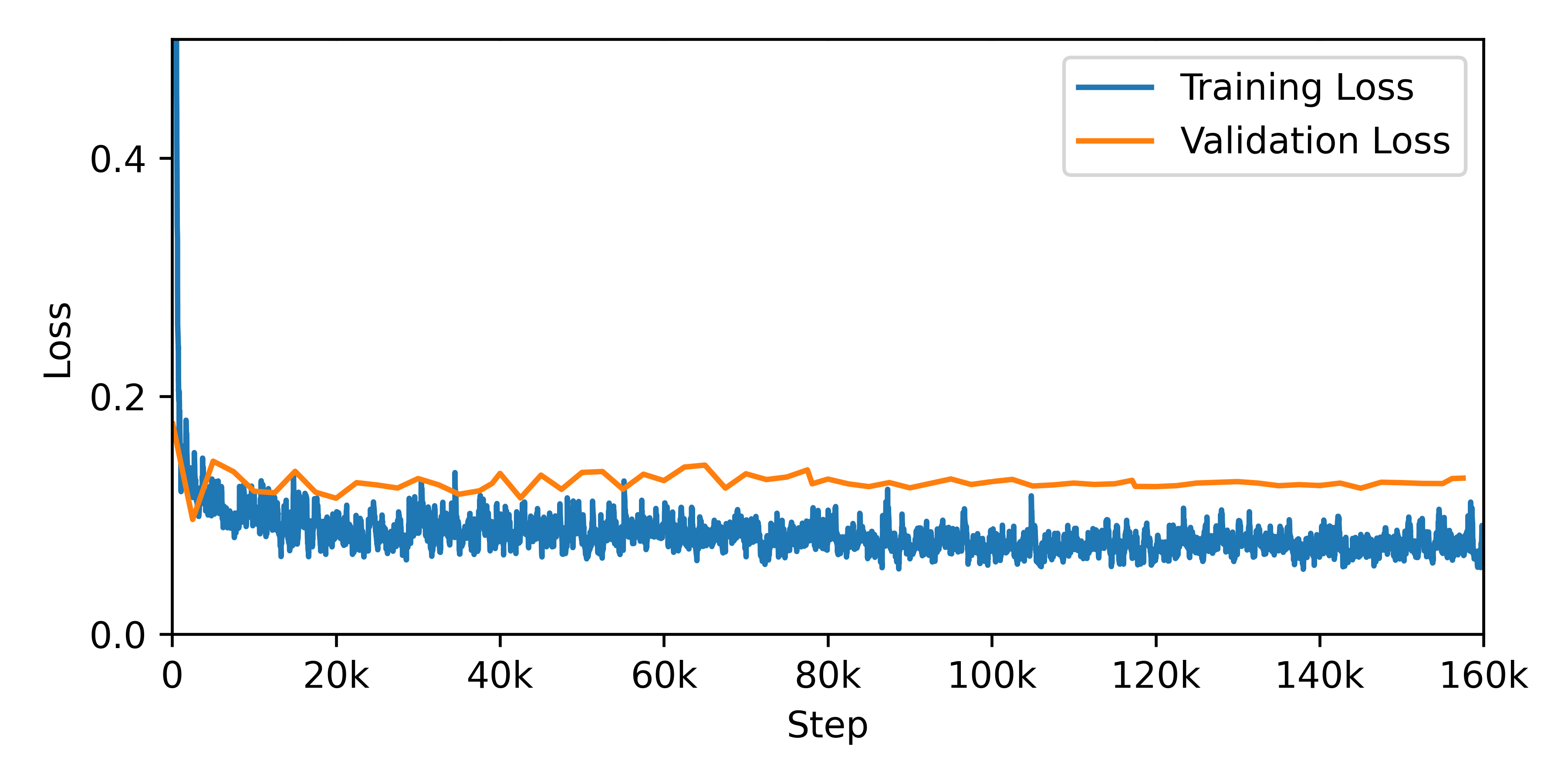}
    \caption{Training and validation loss for the detection network.}
    \label{fig:loss}
\end{figure}

\begin{figure}[h!]
    \centering
    \begin{subfigure}[b]{\linewidth}
        \centering
            \begin{subfigure}[b]{0.49\linewidth}
            \centering
            \includegraphics[width=\linewidth]{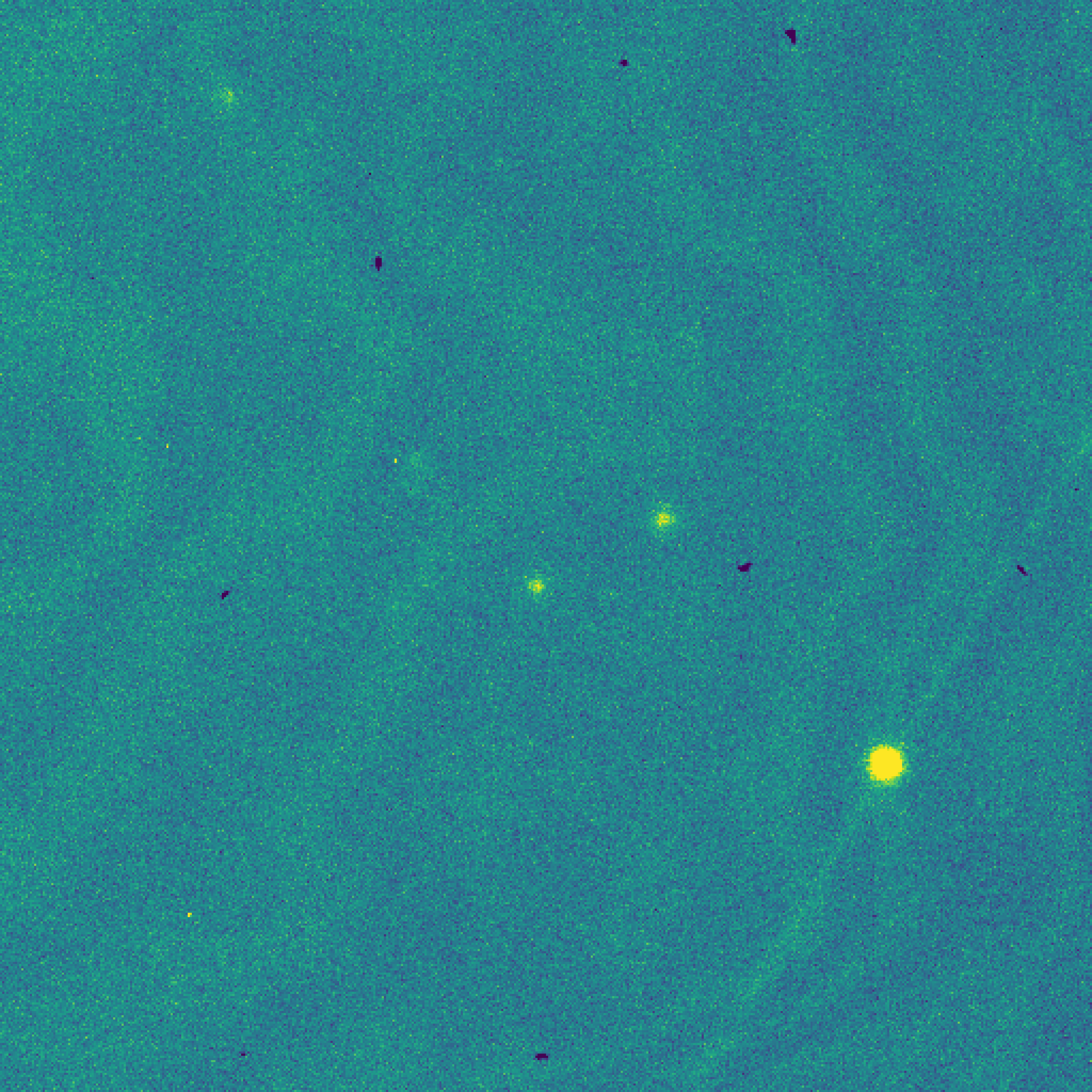}
            \end{subfigure}
            \hfill
            \begin{subfigure}[b]{0.49\linewidth}
            \centering
            \includegraphics[width=\linewidth]{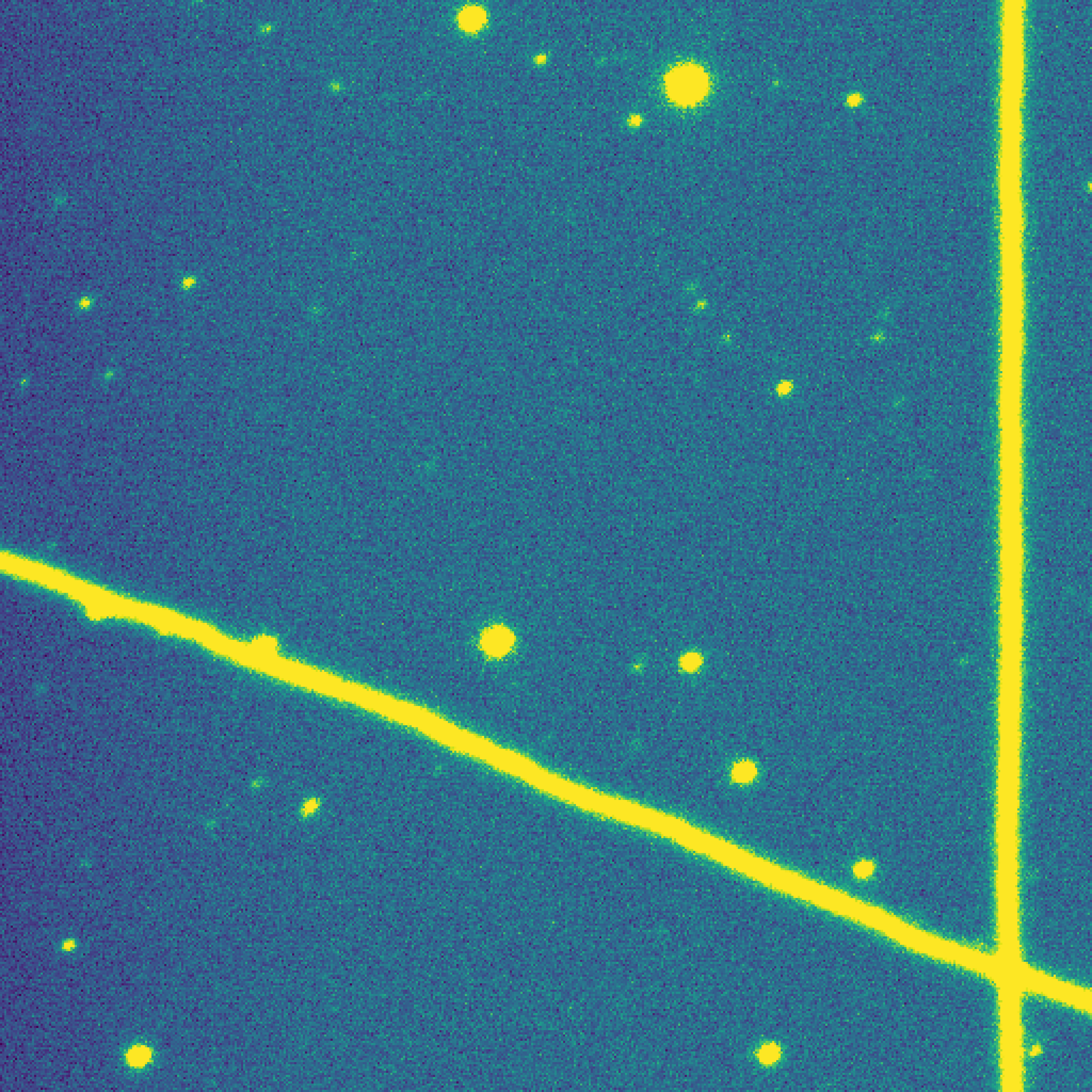}
            \end{subfigure}
        \caption{}
    \end{subfigure}
    \begin{subfigure}[b]{\linewidth}
        \centering
            \begin{subfigure}[b]{0.49\linewidth}
            \centering
            \includegraphics[width=\linewidth]{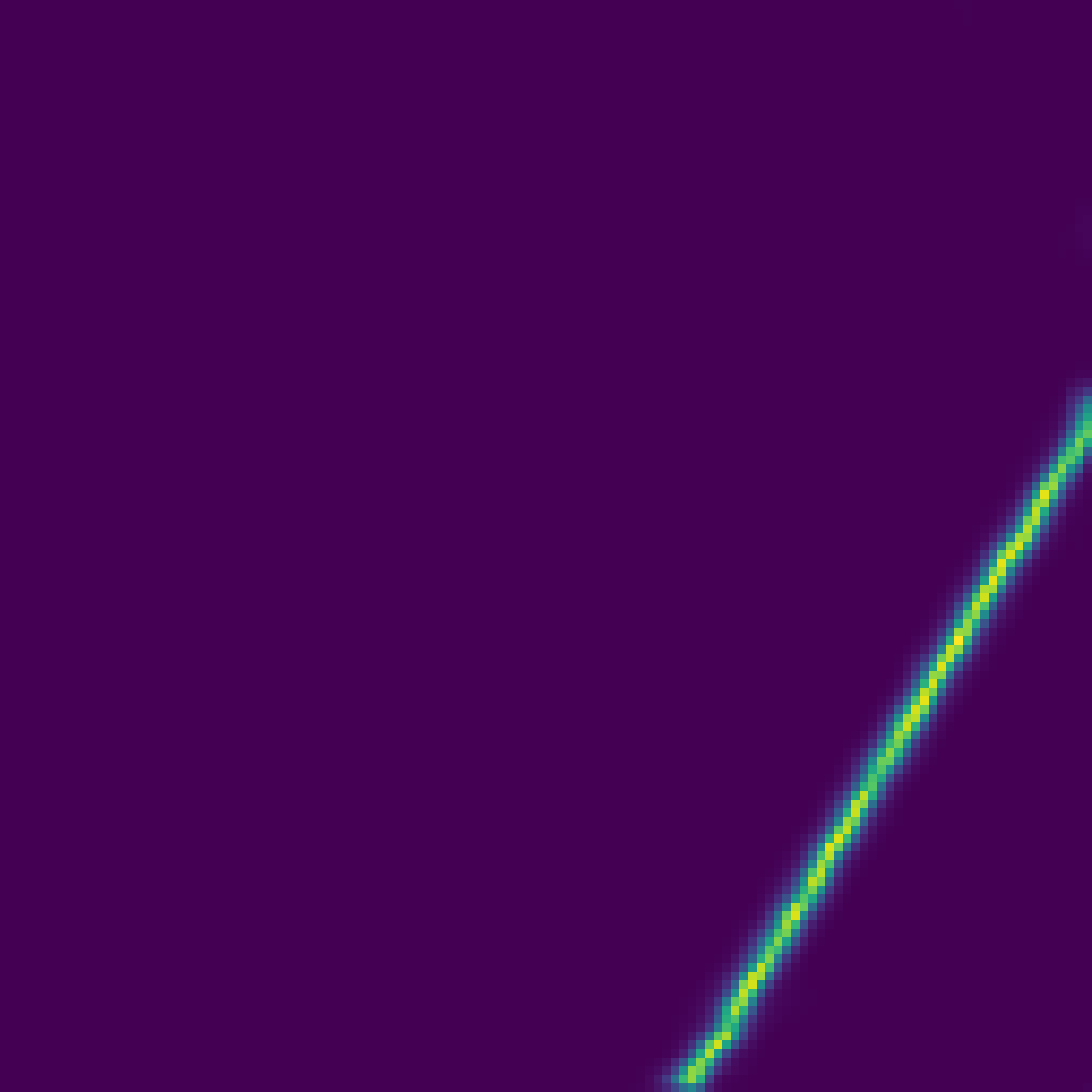}
            \end{subfigure}
            \hfill
            \begin{subfigure}[b]{0.49\linewidth}
            \centering
            \includegraphics[width=\linewidth]{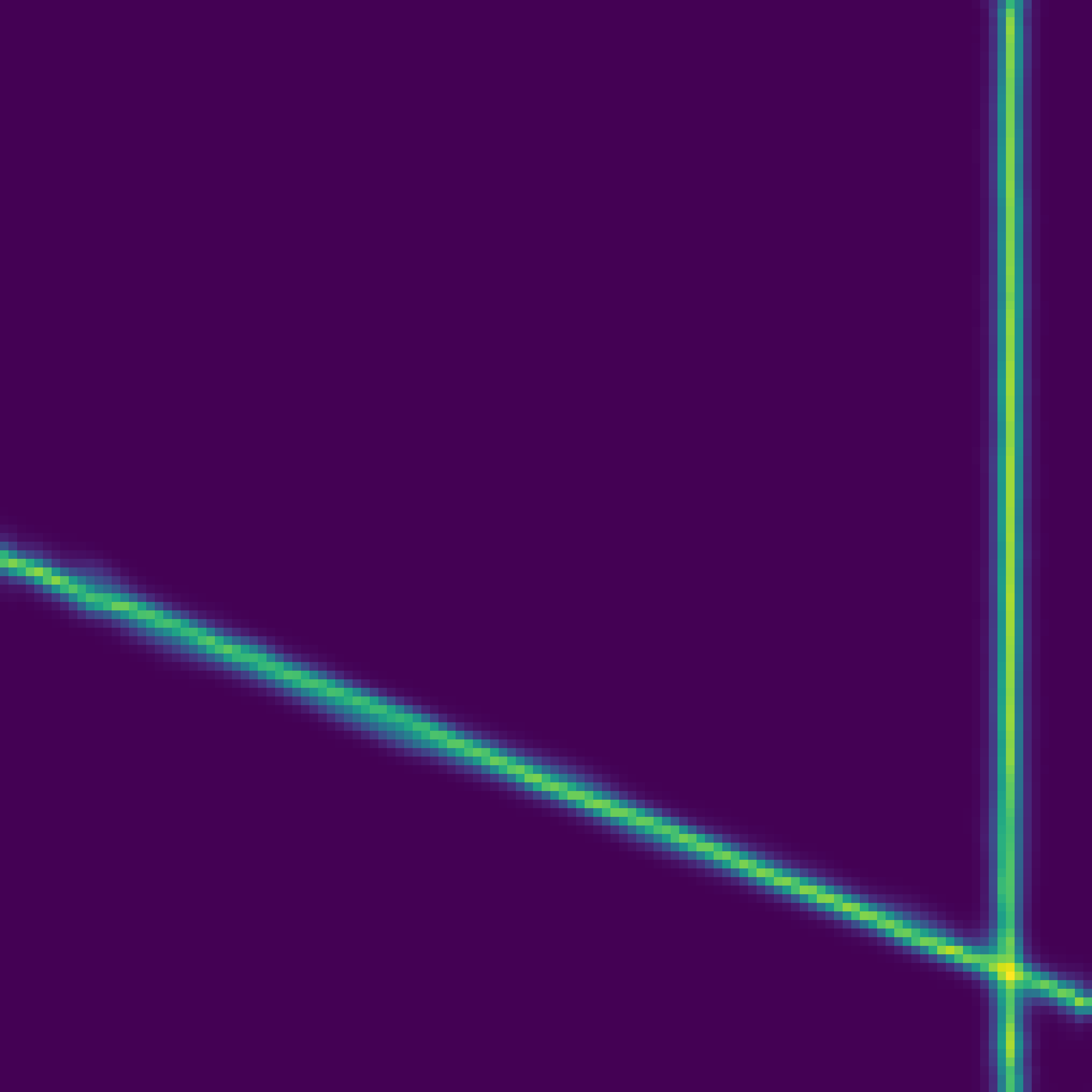}
            \end{subfigure}
        \caption{}
    \end{subfigure}
    \begin{subfigure}[b]{\linewidth}
        \centering
            \begin{subfigure}[b]{0.49\linewidth}
            \centering
            \includegraphics[width=\linewidth]{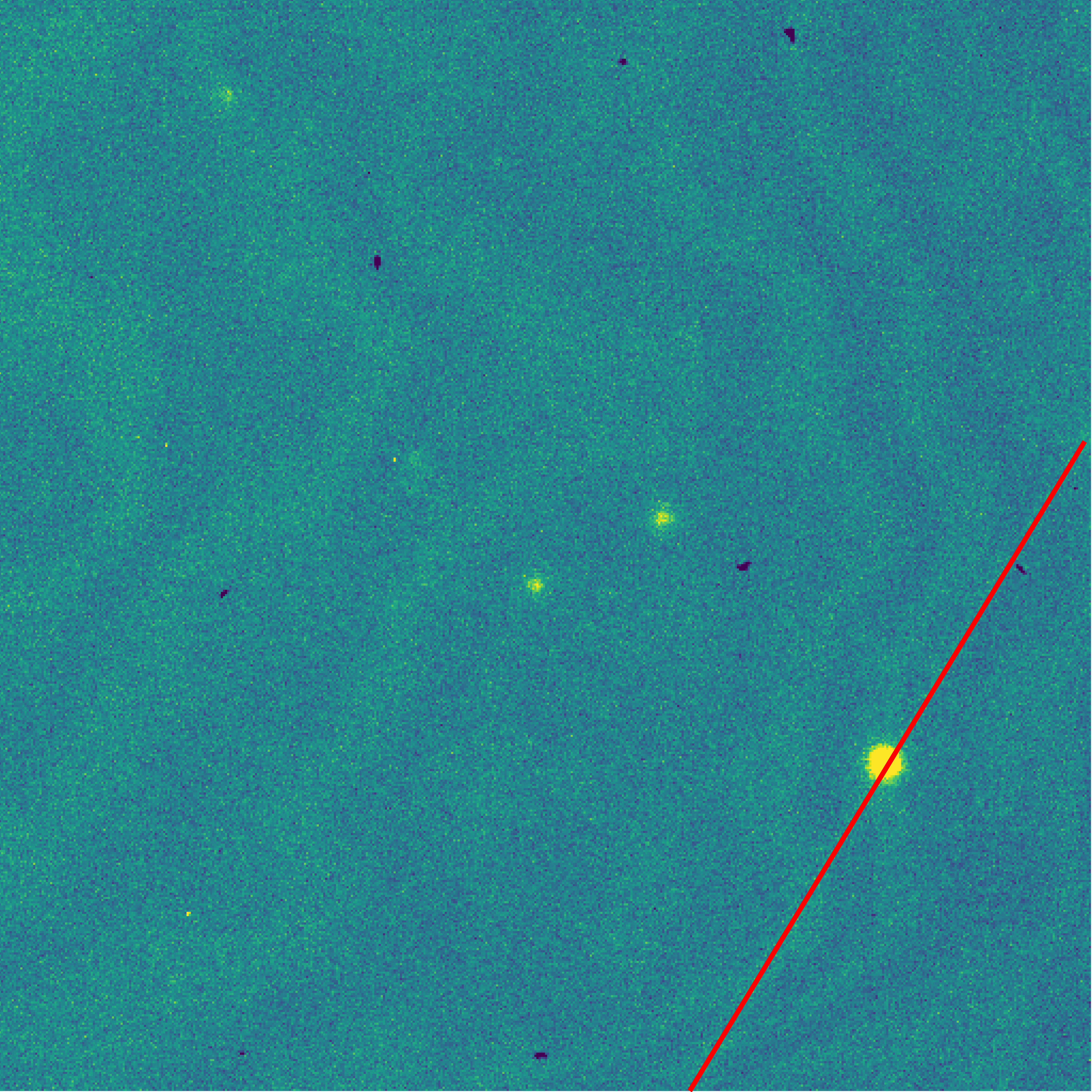}
            \end{subfigure}
            \hfill
            \begin{subfigure}[b]{0.49\linewidth}
            \centering
            \includegraphics[width=\linewidth]{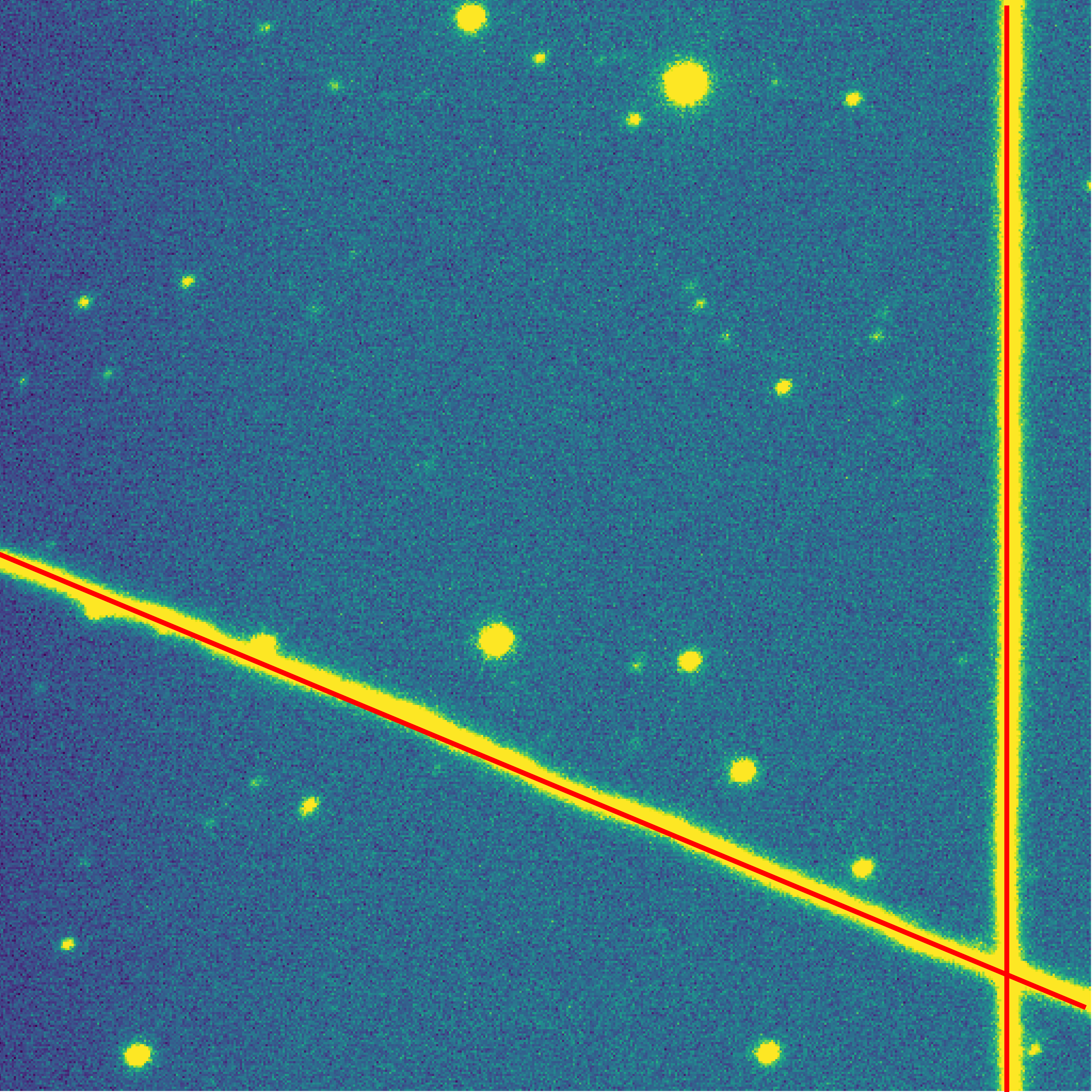}
            \end{subfigure}
        \caption{}
    \end{subfigure}
    \caption{Sample detections with the HT-LCNN network: original image (a), detection heatmap, and detected lines (c). }
    \label{fig:detection_samples}
\end{figure}

Finally, we trained the algorithm using the Adam \citep{kingma_adam_2015} optimizer with an initial learning rate of $1\times 10^{-4}$ and a binary cross-entropy (BCE) loss. To mitigate potential convergence issues, we adopted the AMSGrad variant of Adam \citep{reddi_convergence_2019} and applied L2 regularisation with a strength of $2\times 10^{-5}$. We trained the model over four epochs, leveraging the large dataset size, and scheduled the learning rate to update every two epochs. The progression of training and validation losses throughout the process is shown in Fig.~\ref{fig:loss}. Typical outputs of the algorithm are presented in Fig.~\ref{fig:detection_samples}. The network has two independent output heads: one produces a heatmap that gives, for each pixel, the probability of belonging to a streak, and the other directly predicts the line endpoints. The heatmap is currently used only for visualisation, but could later support the creation of segmentation masks for the detected streaks.

We first evaluated the performance of the neural network on the validation dataset. We assessed the performance using precision, recall, and F1-score. Precision measures the fraction of detections classified as positive that are verified as correct (i.e. true positives) via
\begin{equation}
    Precision = \frac{TP~(\text{True Positive})}{TP + FP~(\text{False Positive})},
\end{equation}
indicating the reliability of the algorithm. Recall measures the fraction of all real (ground-truth) positives that are successfully detected via
\begin{equation}
    Recall = \frac{TP}{TP + FN~(\text{False Negative})},
\end{equation}
reflecting the sensitivity of the algorithm. The F1-score, the harmonic mean of precision and recall via
\begin{equation}
    F1 = 2 \cdot \frac{Precision \cdot Recall}{Precision + Recall},
\end{equation}
provides a single measure of overall performance. A detection is considered to match the ground truth if the corresponding endpoints are within 20 pixels of each other and their orientations differ by no more than ten degrees. To evaluate the impact of our line-stitching procedure, we calculated the metrics at both the crop level and the full CCD level, as can be seen in Table~\ref{tab:performance_validation}. The improved scores after stitching confirm the effectiveness of this method.

\begin{table}[h!]
    \centering
    \caption{Performance of the detection algorithm on the validation dataset.}
    \begin{tabular}{c|c|c|c}
        \hline \hline
        Evaluation level & Precision & Recall & F1-score\\
        \hline
        Crop-wise & 0.944 & 0.939 & 0.941\\
        CCD-wise & 0.971 & 0.961 & 0.966
    \end{tabular}
    \label{tab:performance_validation}
\end{table}

\begin{figure}[h!]
    \centering
    \includegraphics[width=\linewidth]{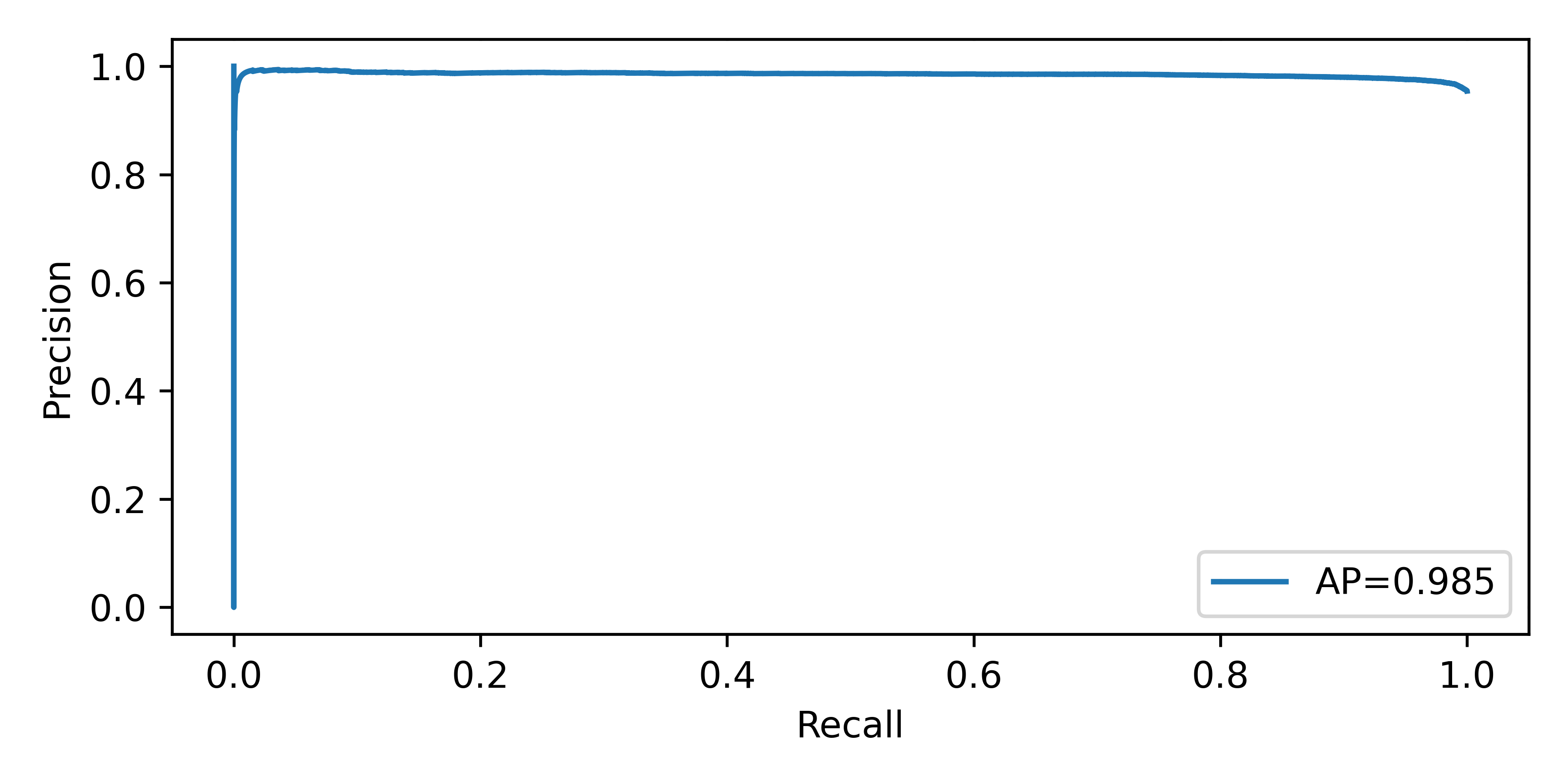}
    \caption{Detection algorithm precision-recall curve, evaluated CCD-wise.}
    \label{fig:pr_curve}
\end{figure}

The algorithm provides a confidence score for each detection, together with the detections themselves. By varying the confidence threshold, we generated a precision–recall curve that shows how threshold selection affects algorithm performance. Both precision and recall remain consistently high and balanced across the full range of thresholds, indicating the robustness of the algorithm (see Fig.~\ref{fig:pr_curve}). We observed no bias towards any over- or under-detection. On the validation set, the algorithm reaches an average precision (area under the precision–recall curve) of 0.985. We explored different detection thresholds to optimise performance. However, given the shape of the precision–recall curve and to avoid missing potential true positives, we did not apply a fixed threshold and kept all candidate detections. We acknowledge that this may result in a higher number of false positives but adopted this intentionally conservative approach to maximise completeness at this stage, relying on our classifier to discard false-positive detections in a subsequent step.

\begin{figure}[h!]
    \centering
    \includegraphics[width=\linewidth]{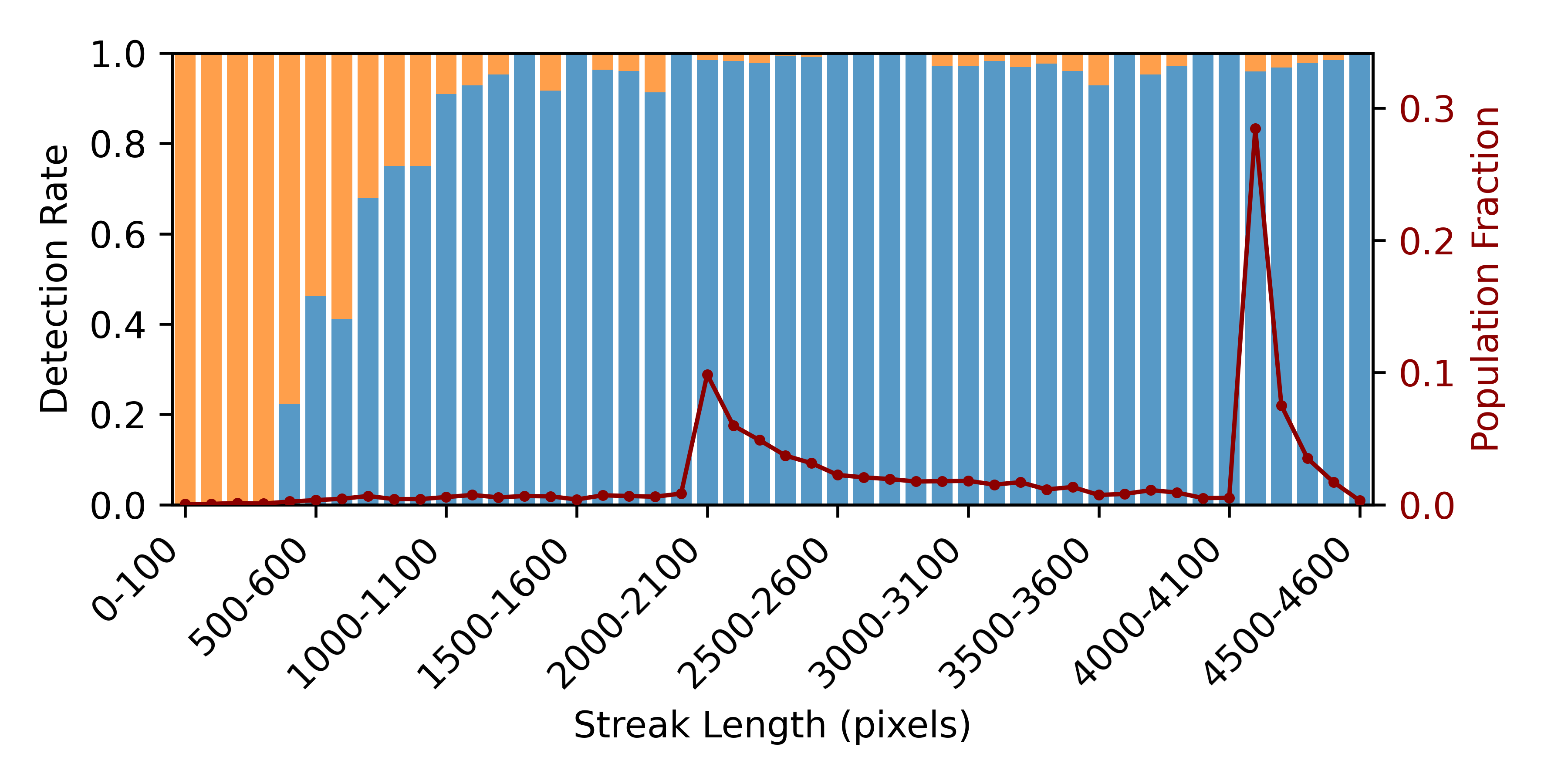}
    \caption{Detection performance as a function of streak length for artificial streaks in the validation dataset (100-pixel-wide bins). Blue bars show the true positive detection rate, orange bars the false negative complement, and the red line the fraction of the total population per bin.}
    \label{fig:length_bins}
\end{figure}

\begin{figure}[h!]
    \centering
    \includegraphics[width=\linewidth]{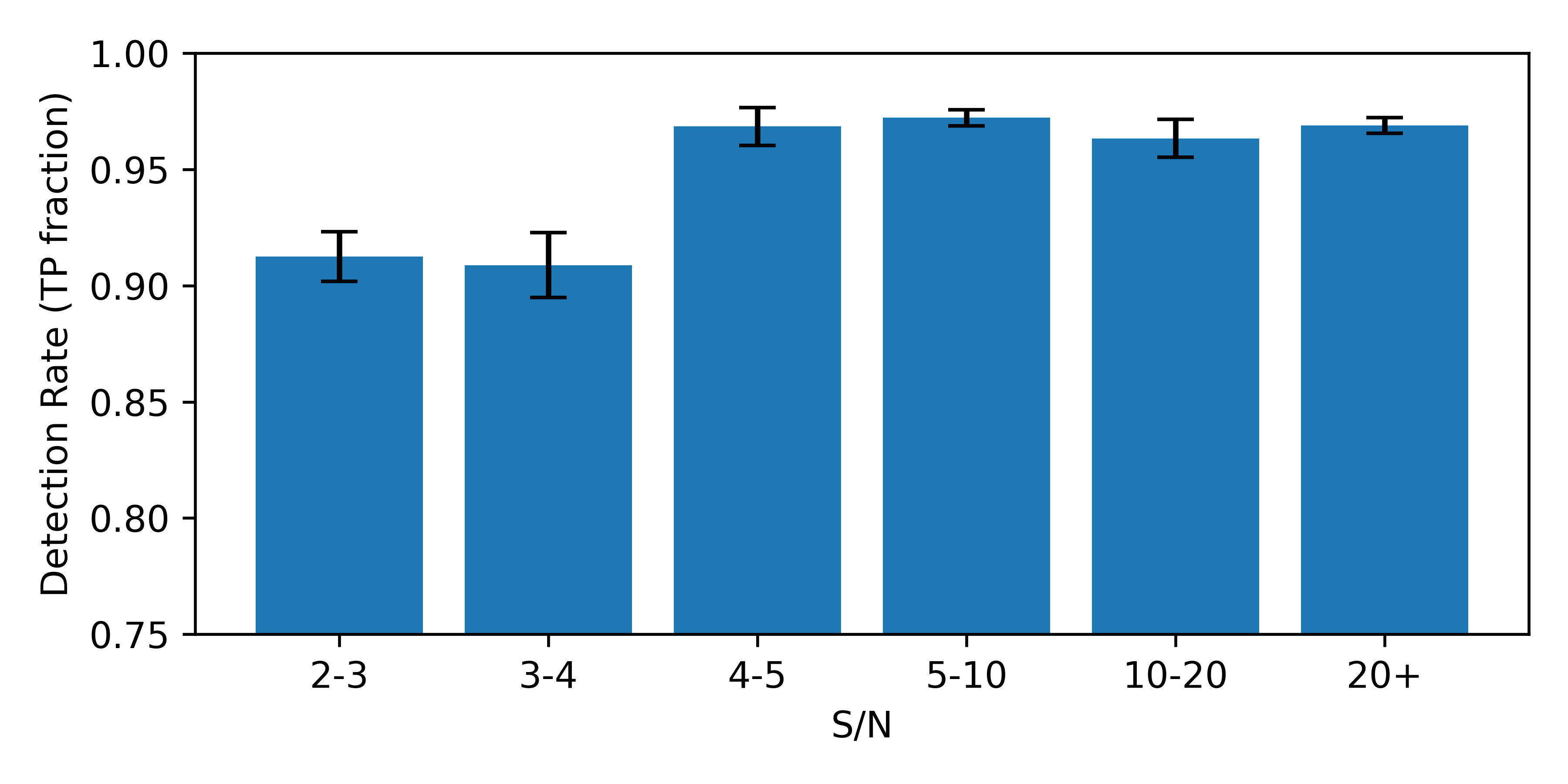}
    \caption{Detection completeness as a function of streak S/N for artificial streaks in the validation set. Error bars indicate the standard deviation estimated from ten bootstrap realisations of the full validation set (sampling with replacement), with larger uncertainties at low S/N reflecting the smaller number of streaks in those bins.}
    \label{fig:snr_bins}
\end{figure}

Figure~\ref{fig:length_bins} reports the artificial streak population distribution across streak length, as well as the detection rate versus streak length. The largest fraction of streaks relative to the total population clusters near the CCD height (4\,102 pixels) and width (2\,048 pixels) dimensions, reflecting our design where 80\% of simulated streaks extend across the full CCD. Streaks longer than 1\,000 pixels were reliably detected. The algorithm performed less well on shorter streaks, which are also under-represented in the training set, and streaks confined to a single 512\,$\times$\,512 pixel crop were automatically discarded by the stitching procedure. In terms of S/N dependence, the algorithm detected more than 95\% of streaks with S/N values greater than four. For streaks with S/N between two and four, the detection performance was more sensitive to background conditions, such as image contamination and field complexity, yet the detection rate typically remained above or around 90\%, as illustrated in Fig.~\ref{fig:snr_bins}.


We performed a final assessment of the detection algorithm on the test dataset, which the network had never seen before. Table~\ref{tab:performance_testset} reports the precision, recall, and F1-score on the test set, evaluated crop-wise and CCD-wise. These values agree with those obtained on the validation set and indicate good generalisation without any obvious biases.

\begin{table}[h!]
    \centering
    \caption{Performance of the detection algorithm on the test dataset.}
    \begin{tabular}{c|c|c|c}
        \hline \hline
        Evaluation level & Precision & Recall & F1-score\\
        \hline
        Crop-wise & 0.944 & 0.929 & 0.936\\
        CCD-wise & 0.962 & 0.953 & 0.958
    \end{tabular}
    \label{tab:performance_testset}
\end{table}

\subsection{Generalisation to real world data}\label{sec:generalisation}
While the validation and test results provide useful insights, they do not fully capture the algorithm’s performance on real data, especially given the large fraction of artificial streaks in the dataset. To obtain a more robust characterisation of the model’s behaviour under real-world conditions, we applied the detection pipeline to all OmegaCAM observations from December 2023, consisting of 2\,357 full-mosaic images. Given the size of this dataset, full annotation was not feasible, so we estimated the false-positive rate by manually inspecting the individual CCD-level detections and discarding erroneous cases. This procedure allowed us to estimate the precision of the network but not the recall, which would require complete annotation of the dataset.

From the 3\,370 detected streaks (CCD-level), this large-scale evaluation resulted in a precision of 0.783, about 18\% lower than on the test set. The most common sources of false positives were diffraction spikes from bright stars, saturation bleeding from overexposed sources, fringe patterns due to imperfect filters, and stray light. This highlights that a dedicated classifier to reject false positives is essential to further improve the precision of the pipeline.

\subsection{Training and performance of the classifier}
We trained the VGG6 classification algorithm using the AdamW optimiser and a BCE loss with an initial learning rate of $4.5432\times 10^{-4}$ and a weight decay of $6.5207\times 10^{-5}$. The network was trained for 100 epochs with a batch size of 32.

\begin{figure}[h!]
    \centering
    \includegraphics[width=\linewidth]{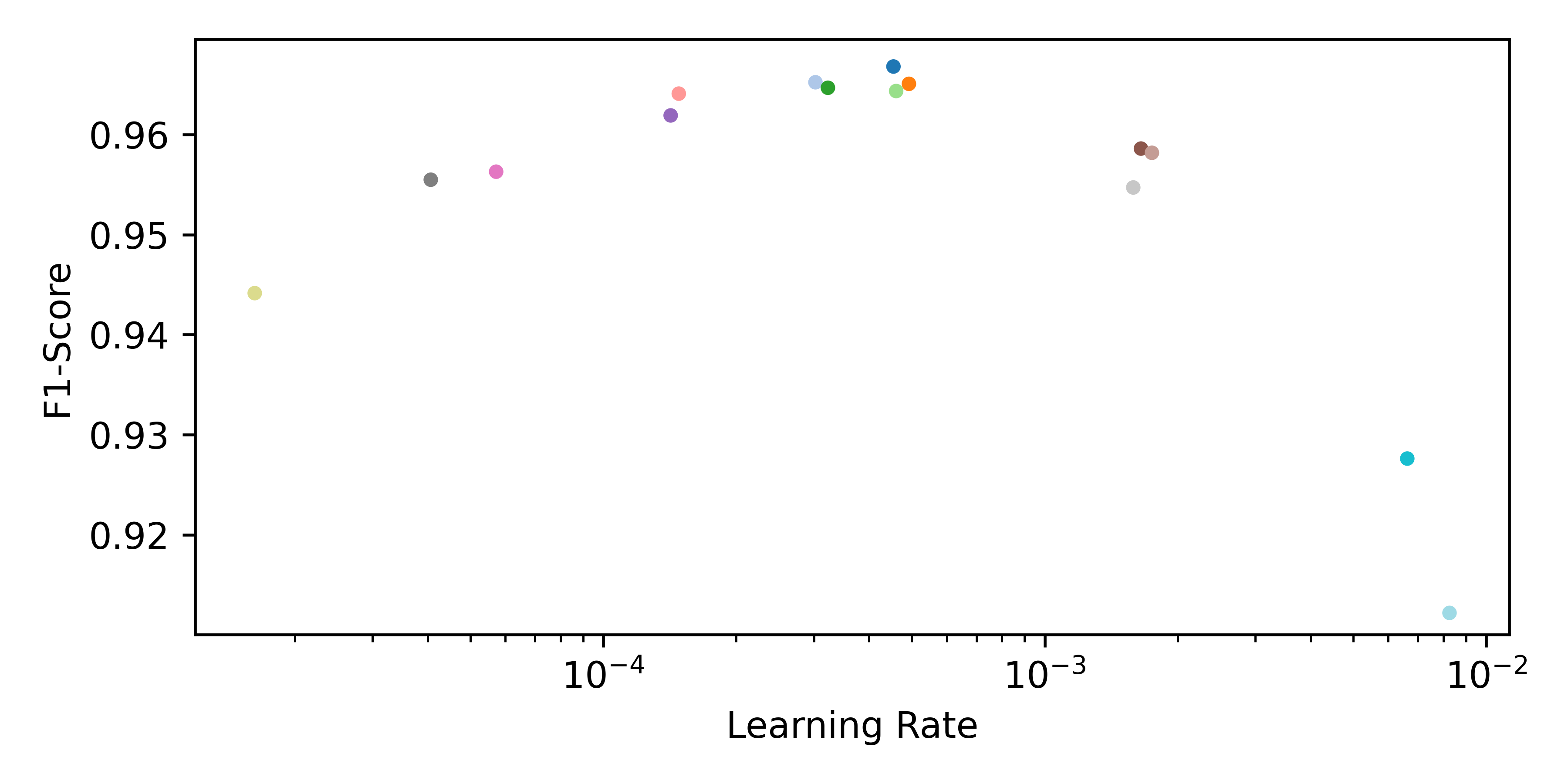}
    \caption{Classifier F1-score on the validation set as a function of the learning rate.}
    \label{fig:classifier_lr_search}
\end{figure}

\begin{figure}[h!]
    \centering
    \includegraphics[width=\linewidth]{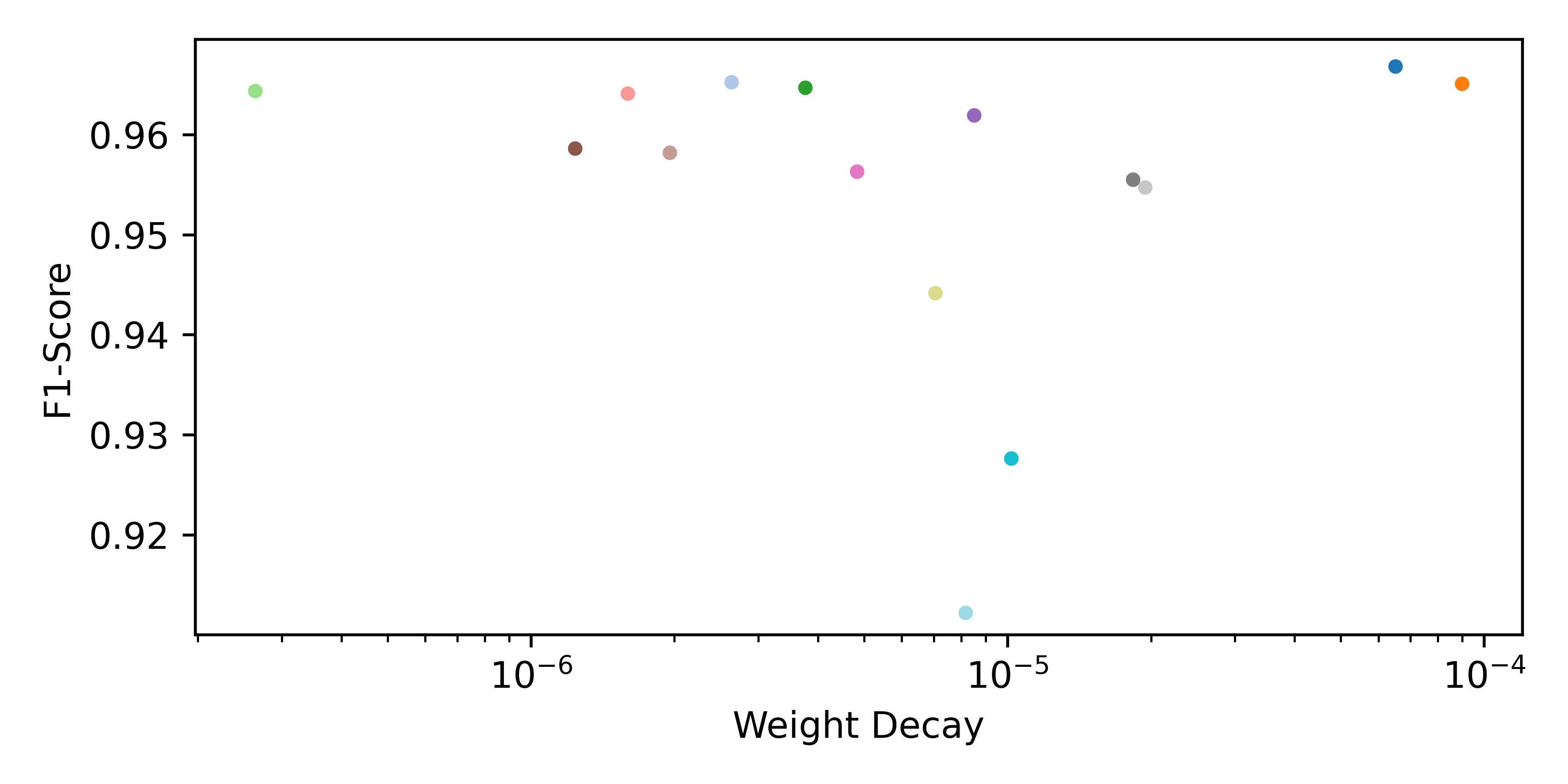}
    \caption{Classifier F1-score on the validation set as a function of the weight decay.}
    \label{fig:classifier_wd_search}
\end{figure}

We determined the initial learning rate by training the model with several values randomly selected between $1\times 10^{-2}$ and $1\times 10^{-5}$. The learning rate that produced the highest F1-score on the validation set was chosen. The F1-scores corresponding to each tested learning rate are shown in Fig.~\ref{fig:classifier_lr_search}. We applied the same procedure to identify the optimal weight decay, as illustrated in Fig.~\ref{fig:classifier_wd_search}.

\begin{figure}[t]
    \centering
    \includegraphics[width=\linewidth]{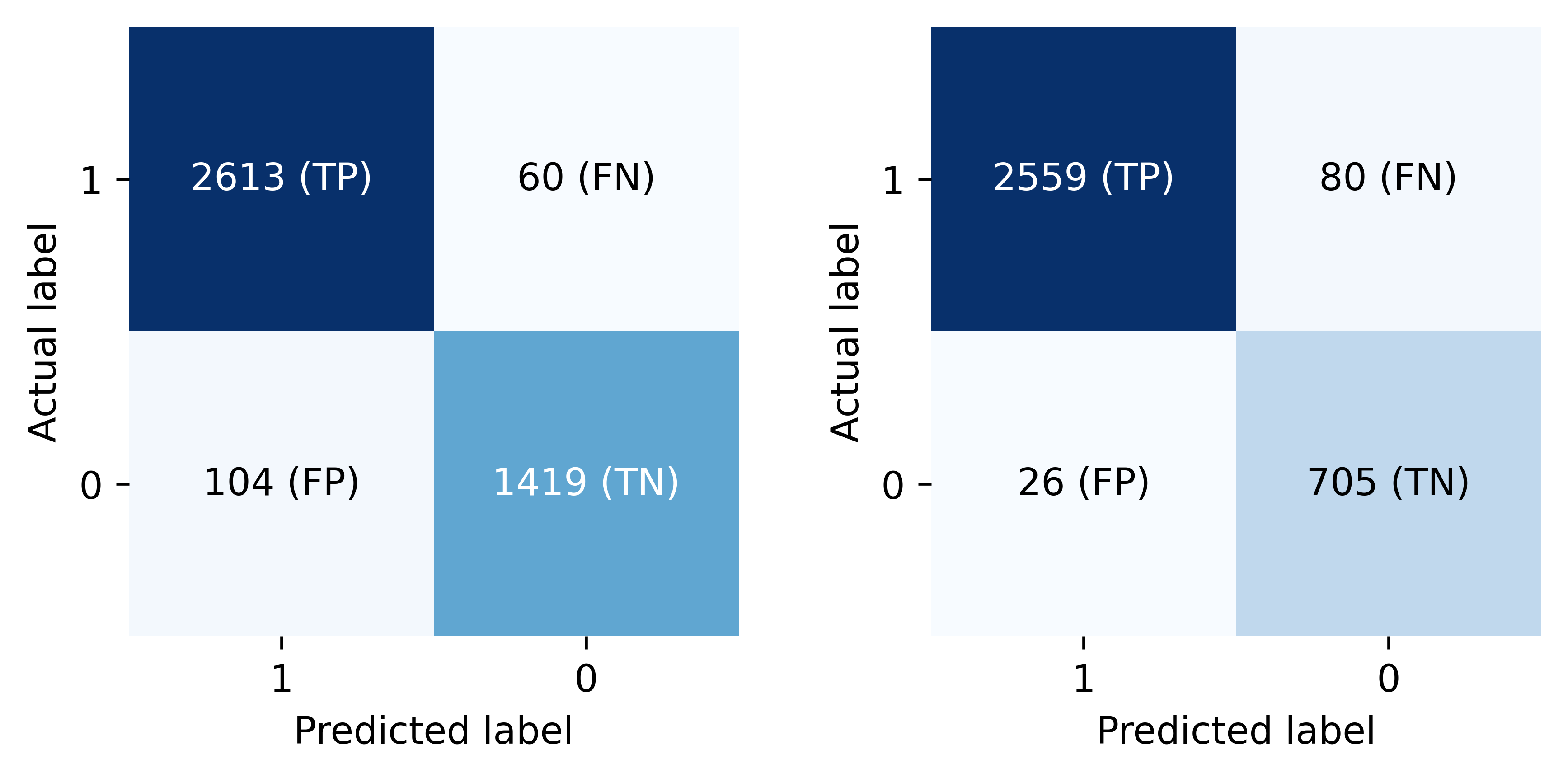}
    \caption{Classifier confusion matrices on the validation (left) and test (right) datasets, with label 1 indicating real streaks and label 0 indicating non-streak detections.}
    \label{fig:cm}
\end{figure}

\begin{figure}[t]
    \centering
    \includegraphics[width=\linewidth]{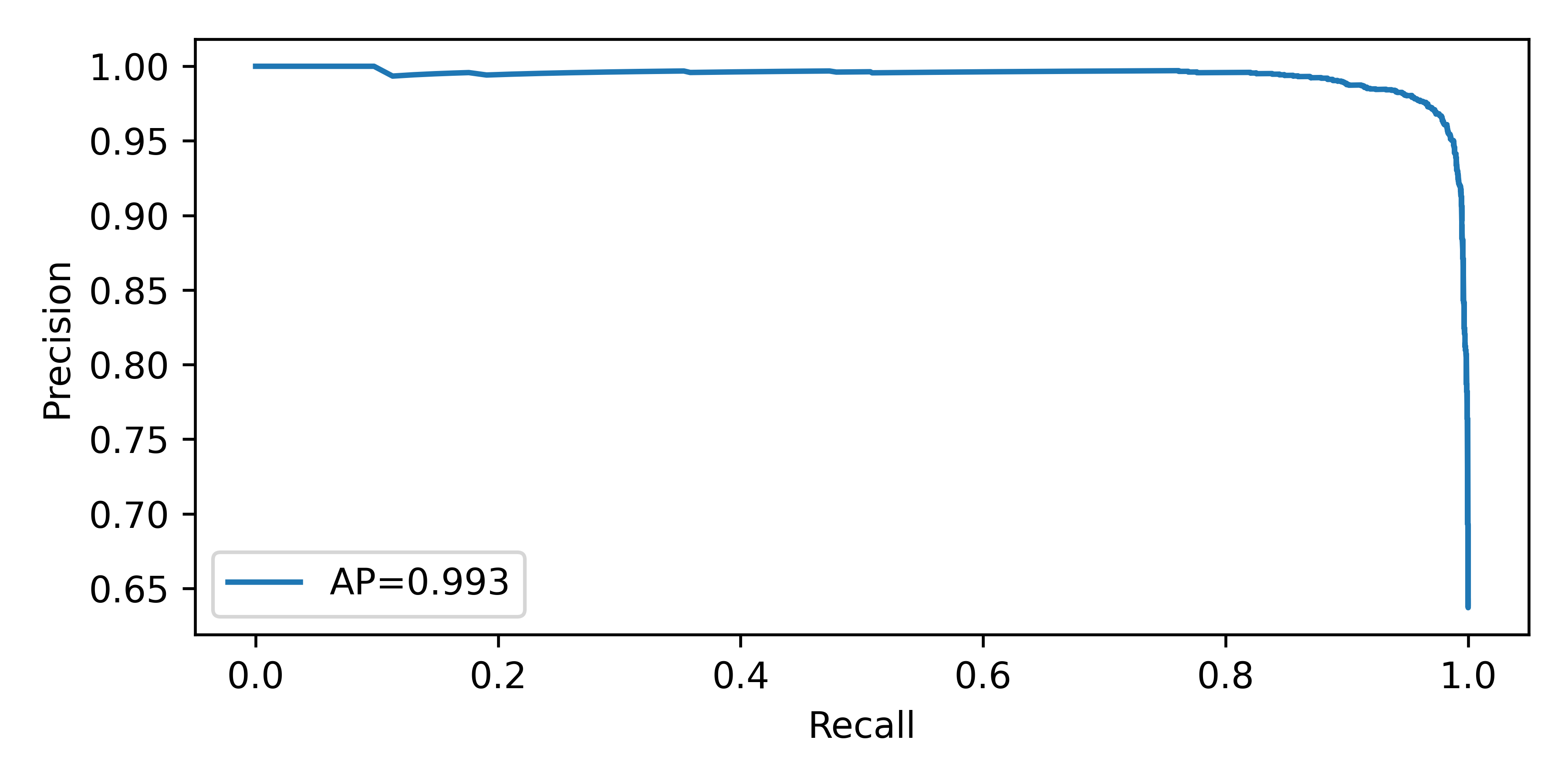}
    \caption{Classifier precision-recall curve.}
    \label{fig:pr_curve_classifier}
\end{figure}

As with the detection algorithm, we first evaluated the final network on the validation dataset. Table~\ref{tab:performance_classifier} lists the results in terms of precision, recall, and F1-score. The corresponding confusion matrix in Fig.~\ref{fig:cm} includes the true negative classifications, providing a more complete picture of performance. The precision–recall curve in Fig.~\ref{fig:pr_curve_classifier} shows no visible bias and confirms the network’s strong results. Tuning the decision threshold did not yield a significant improvement, thus we kept the standard threshold of 0.5, classifying outputs above 0.5 as true positives and those below as false negatives.

\begin{figure}[h!]
    \centering
    \includegraphics[width=\linewidth]{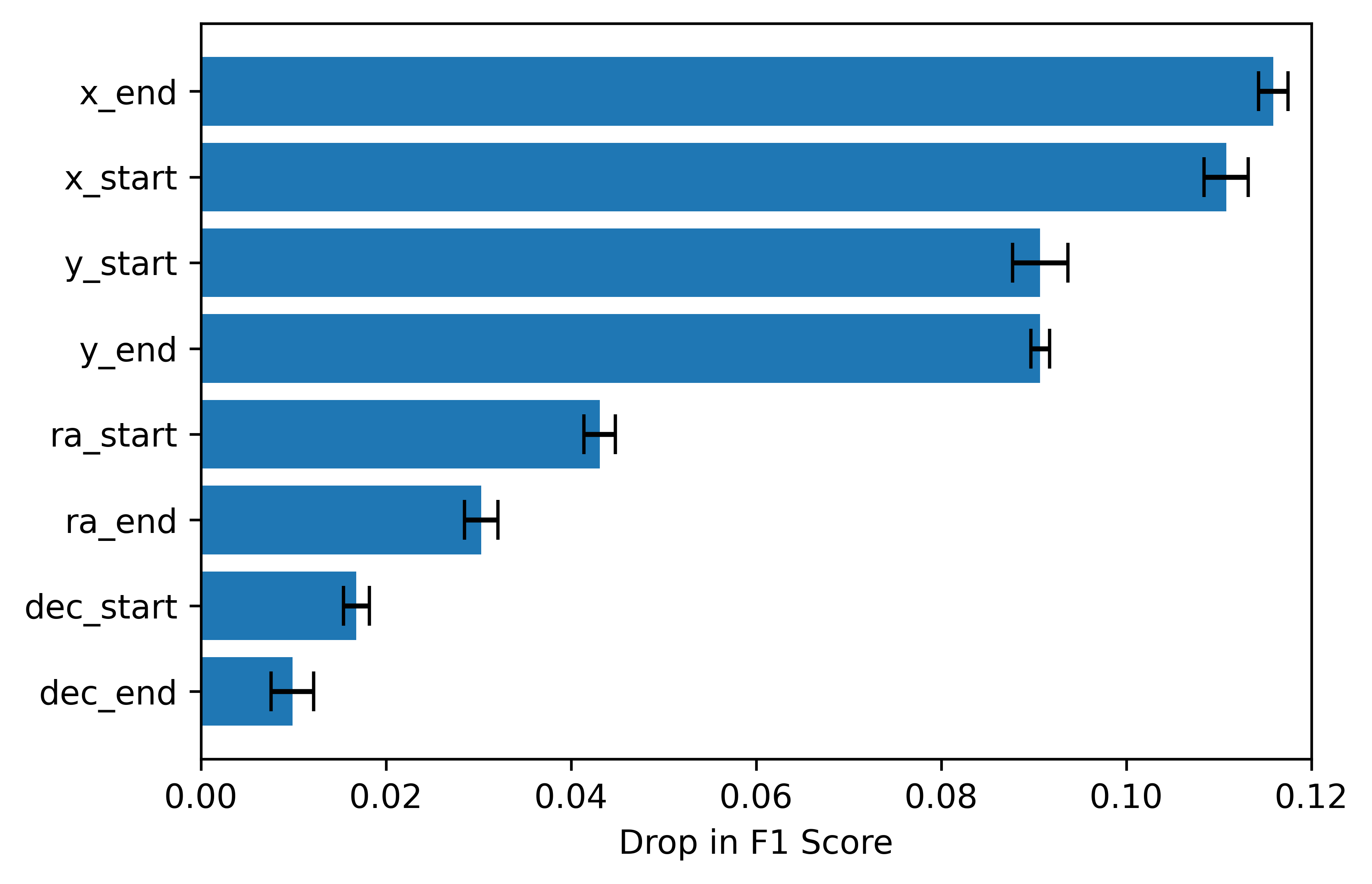}
    \caption{Permutation-based feature importance for the MLP. Bars show the average drop in validation F1-score when each input feature is randomly permuted, with error bars computed over multiple permutations.}
    \label{fig:mlp_importance}
\end{figure}

For comparison, the coordinate-only MLP described in Sect.~\ref{sec:methods-classifier} achieves an F1-score of 0.882 on the same validation set, substantially below the image-based VGG6 classifier (0.970). We also assessed the relative importance of the input features for the MLP by randomly permuting them one at a time. As shown in Fig.~\ref{fig:mlp_importance}, the X and Y image coordinates have the greatest influence on the network’s performance, while the sky coordinates have a smaller contribution. A manual inspection of the discarded detections suggests that the MLP has a strong tendency to retain streaks that cross the full detector while rejecting detections that do not span the entire CCD. Although this behaviour helps remove many false positives, it also eliminates a number of streaks with an endpoint inside the image, which are particularly valuable for the photometric analysis planned at a later stage. Since we cannot afford to systematically lose such detections, the CNN classifier was retained as our final classification method.

\begin{table}[t]
    \centering
    \caption{Performance of the classification algorithm on the validation and test datasets.}
    \begin{tabular}{c|c|c|c}
        \hline \hline
        Dataset & Precision & Recall & F1-score\\
        \hline
        Validation & 0.962 & 0.978 & 0.970\\
        Test & 0.990 & 0.970 & 0.980
    \end{tabular}
    \label{tab:performance_classifier}
\end{table}

Finally, we applied the VGG6 algorithm to the independent test set, which was not used at any stage of training or hyperparameter tuning. Table~\ref{tab:performance_classifier} reports the corresponding precision, recall, and F1-score, and the associated confusion matrix is shown in Fig.~\ref{fig:cm}. The metrics on the test data show a slight improvement compared to those obtained on the validation set.

As the December 2023 data served both to assess the generalisation of the detection algorithm and to provide the test set for the classification algorithm, this dataset enabled an evaluation of the full pipeline's performance (i.e. detection followed by classification). To quantify the improvement brought by the classifier, the false-positive rate was introduced as an additional metric via
\begin{equation}
    \text{FPR} = \frac{FP}{FP + TN~(\text{True Negative})},
\end{equation}
alongside the standard scores reported in Table~\ref{tab:performance_delta}. For this dataset, only the precision of the detection network can be computed directly; the other detection-only metrics were approximated and are used solely to quantify the effect of adding the classifier to the full pipeline. Overall, the classifier corrects 96.4\% of the detection algorithm’s errors while preserving 97.0\% of the true detections. Typical inference times for the detection and classification networks are given in Appendix~\ref{sec:app_time}.

\begin{table}[h!]
    \centering
    \caption{Performance of the streak detection pipeline on the December 2023 data, before and after adding the classifier.}
    \begin{tabular}{c|c|c|c}
        \hline \hline
        Metric & Detection & Detection + classification & $\Delta$\\
        \hline
        Precision & 0.783 & 0.990 & \textcolor{Green3}{$+$0.207}\\
        \midrule
        Recall & \textcolor{gray}{1.000} & 0.970 & \textcolor{red}{$-$0.030}\\
        F1-score & \textcolor{gray}{0.878} & 0.980 & \textcolor{Green3}{$+$0.102}\\
        FPR & \textcolor{gray}{1.000} & 0.036 & \textcolor{Green3}{$-$0.964}
    \end{tabular}
    \tablefoot{Desirable improvements appear in green and unfavourable changes in red. The first row reports measured performance, whereas the remaining rows report estimated values based on placeholder entries (shown in grey) used to assess performance deltas, as the corresponding detection-only values are not available for this dataset.}
    \label{tab:performance_delta}
\end{table}

Because the annotations of the classification datasets also record the type of false positive produced by the detection algorithm, we can assess how effectively the classifier rejects false positives for each false-positive class, as reported in Table~\ref{tab:performance_class}.

\begin{table}[h!]
    \centering
    \caption{Performance of the classifier on different false-positive classes in the December 2023 data.}
    \begin{tabular}{c|c|c|c}
        \hline \hline
        FP class & Total & Correctly classified & Rejection Rate\\
        \hline
        Fringes & 356 & 352 & 98.88\%\\
        Diffraction & 165 & 160 & 96.97\%\\
        Saturation & 134 & 130 & 97.01\%\\
        Misc. & 76 & 63 & 82.89\%
    \end{tabular}
    \label{tab:performance_class}
\end{table}

\subsection{Application to OmegaCAM data}
The manual review of all detections from December 2023, as described in Sect.~\ref{sec:generalisation}, allowed us to evaluate the fraction of genuine detections that could not be matched to any object in the space-track catalogue, as shown in Table~\ref{tab:correlation_dec}. The results indicate that approximately 21\% of the confirmed detections did not match any known satellite or debris. A similar analysis for January 2023 revealed that 23\% of true detections could not be correlated with objects in the space-track catalogue. Taking into account natural monthly fluctuations, we estimate that at least 20\% of true streaks identified in the OmegaCAM archival data originate from uncatalogued or unidentified objects. This finding aligns with results reported by other authors for different datasets \citep{danarianto_prototype_2022, stoppa_automated_2024}.

\begin{table}[h!]
    \centering
    \caption{Correlation of December 2023 detections with the space-track catalogue.}
    \begin{tabular}{c|c|c}
    \hline \hline
        True positives & Correlated & Uncorrelated \\
        \hline
        2639 & 2081 & 558
    \end{tabular}
    \label{tab:correlation_dec}
\end{table}

To estimate the number and base characteristics of streaks present in the OmegaCAM archive, we applied our detection algorithm and classifier to all data from 2023. The reported statistics should not be considered fully independent because this dataset was partially used during the pipeline's development; however, the statistics remain indicative of the trends expected in the archive.

\begin{figure*}[h!]
    \centering
    \includegraphics[width=\textwidth]{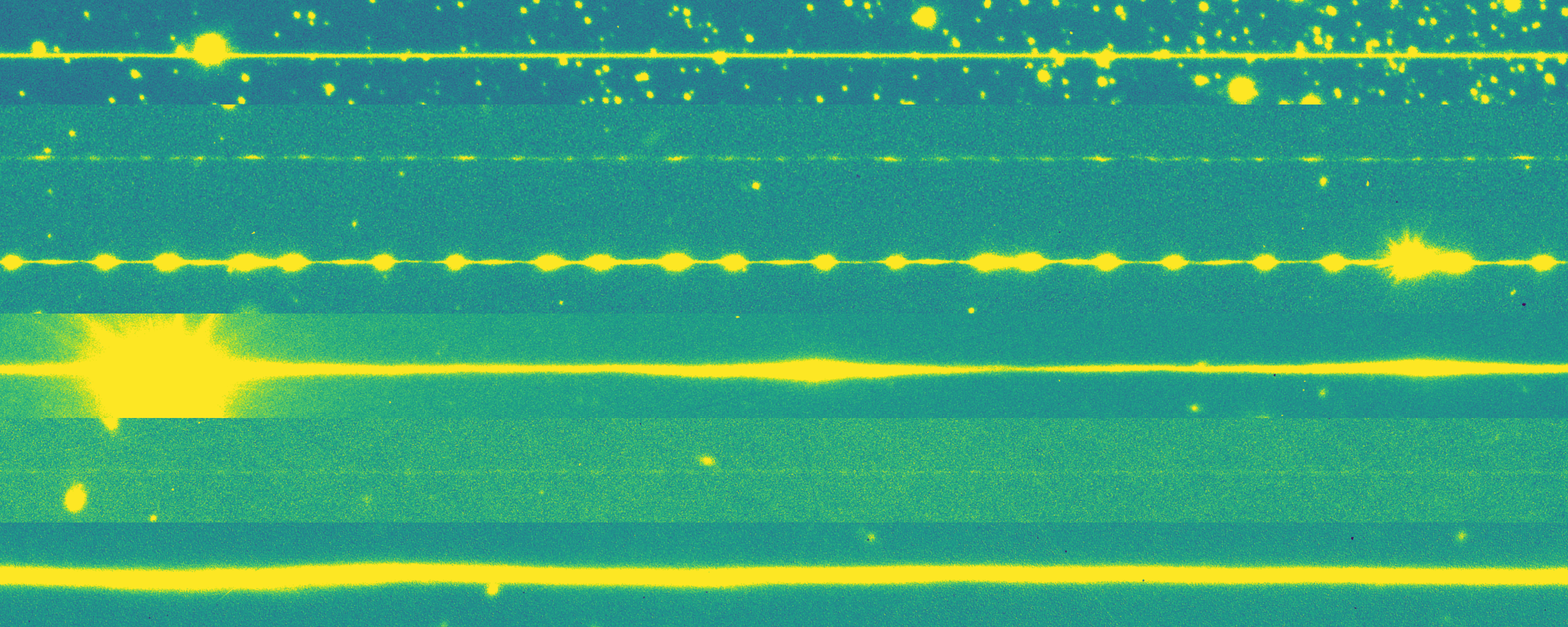}
    \caption{Sample detections from the OmegaCAM archive. All streaks were correlated with known objects.}
    \label{fig:streak_diversity}
\end{figure*}

We analysed a total of 38\,939 OmegaCAM mosaics, corresponding to 1\,246\,048 individual 2\,048\,$\times$\,4\,102 pixel CCDs. Unless explicitly stated otherwise, the number of detections refers to CCD-level detections, meaning that multiple detections within a single OmegaCAM mosaic can belong to the same object. At least one streak appeared in 6\,581 mosaics, or 23\,769 CCDs. Across these affected CCDs, we detected 25\,335 streaks in total. We correlated 16\,776 of them with 2\,761 unique known objects listed in the space-track catalogue, while 8\,559 traces originated from unidentified sources. Several examples of detected streaks are shown in Fig.~\ref{fig:streak_diversity}, illustrating the remarkable diversity of space object traces present in the OmegaCAM data.

\begin{figure*}
    \centering
    \includegraphics[width=\textwidth]{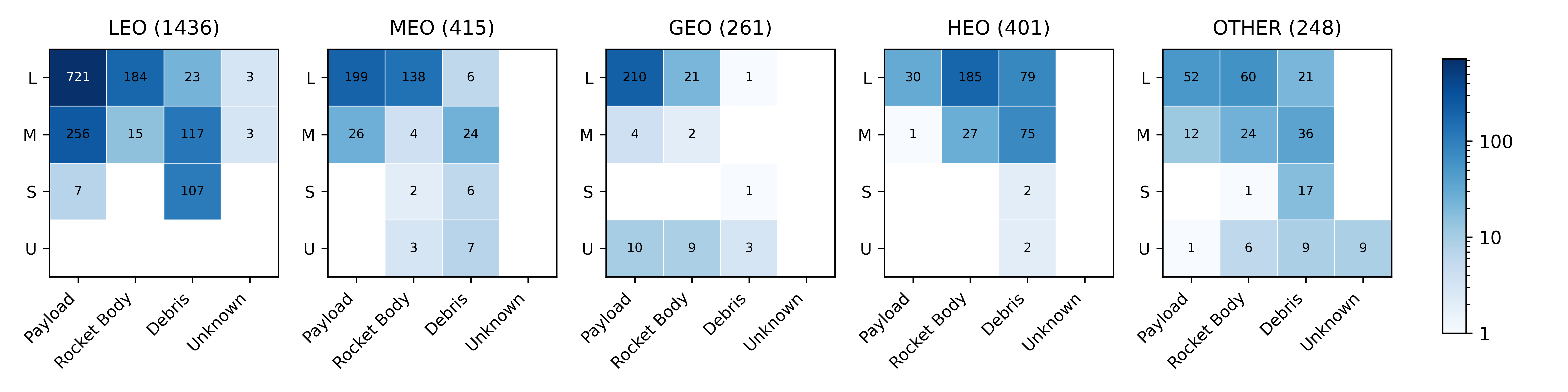}
    \caption{Distribution of correlated objects in 2023 by size and object type for each orbital regime. The number in parentheses next to each orbit label indicates the total number of objects identified in that regime. Object types are shown on the X-axis (payload, rocket body, debris, unknown), while the Y-axis represents RCS sizes: large (L), medium (M), small (S), and unknown (U).}
    \tablefoot{Orbit types: LEO (low-Earth orbit; $<$ 2\,000 km), MEO (medium-Earth orbit; 2\,000 – 35\,700 km), GEO (geosynchronous orbit; 35\,000 - 36\,500 km), HEO (highly eccentric orbit; e $>$ 0.5). RCS sizes: Large ($\geq$ 1 m$^2$), Medium (0.1 – 1 m$^2$), Small ($<$ 0.1 m$^2$).}
    \label{fig:facet_grid_detections}
\end{figure*}

\begin{figure*}
    \centering
    \includegraphics[width=\textwidth]{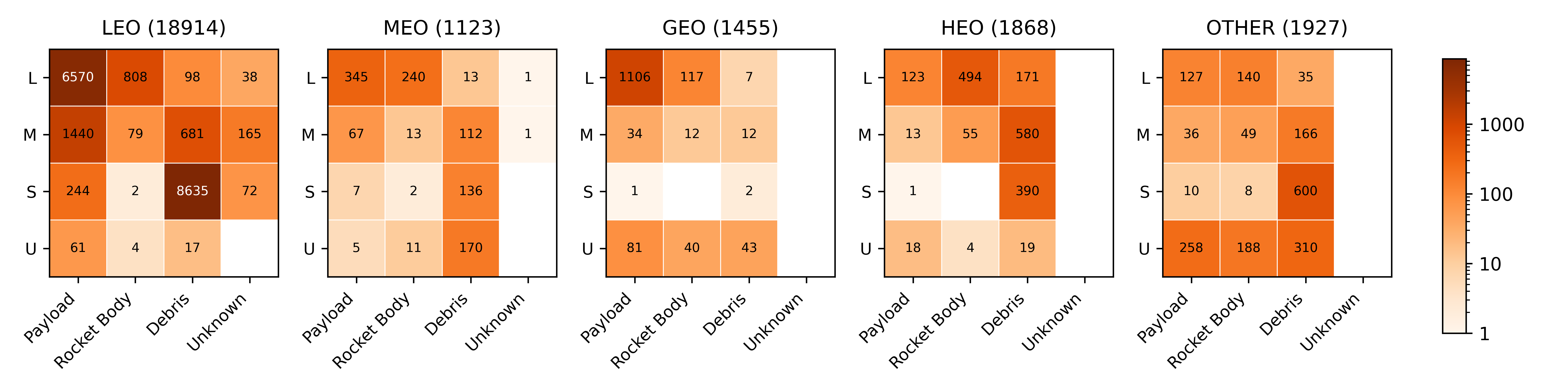}
    \caption{Same as Fig.~\ref{fig:facet_grid_detections} but for all space-track catalogued objects on orbit on 31 December 2023.}
    \label{fig:facet_grid_satcat}
\end{figure*}

\begin{figure*}
    \centering
    \includegraphics[width=\textwidth]{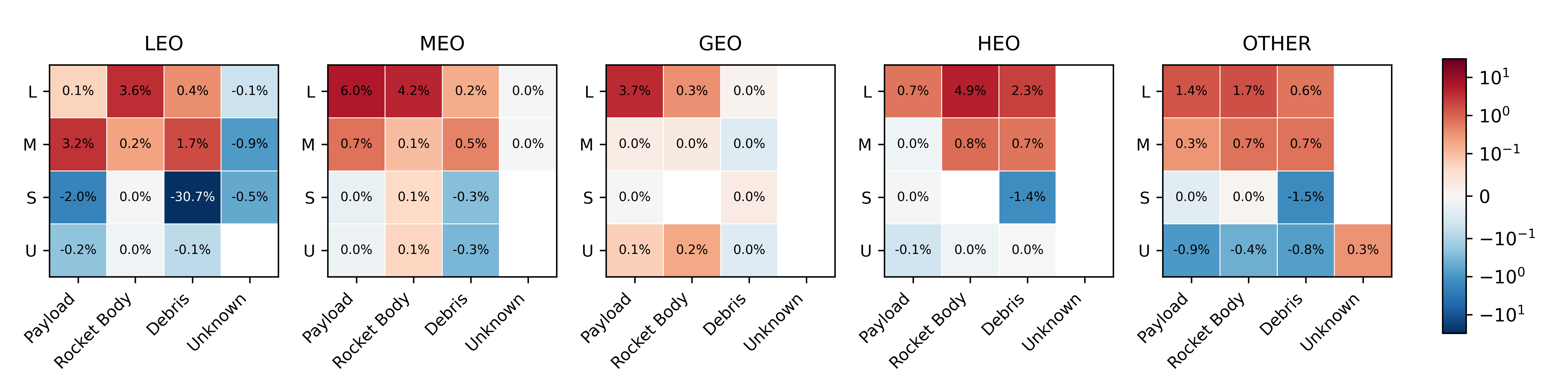}
    \caption{Relative difference in detection ratios between the OmegaCAM 2023 detections and the space-track catalogue. Categories are the same as in Fig.~\ref{fig:facet_grid_detections} and \ref{fig:facet_grid_satcat}.}
    \label{fig:facet_grid_ratio_diff}
\end{figure*}

When we examine the correlated streaks, for which basic information is available, we gain deeper insight into the population distribution, particularly the distribution of detected streaks across different orbital regimes, radar cross-section (RCS) sizes, and object types, as shown in Fig.~\ref{fig:facet_grid_detections}. A comparable plot (Fig.~\ref{fig:facet_grid_satcat}) can be produced for the complete space-track catalogue population that was in orbit on 31 December 2023. By comparing the ratios between these two plots (Fig.~\ref{fig:facet_grid_ratio_diff}), we obtain a clearer view of how effectively we detect objects across orbital regimes, object types, and sizes, and whether any detection biases are evident.

The most striking difference appears in our capability to detect small debris. This is expected, as the space-track catalogue is built using both optical telescopes and radar systems, the latter being particularly effective for detecting small objects in low Earth orbit. Radars are known to be far more sensitive than optical instruments for such targets, so it is unsurprising that OmegaCAM on the VST under-detects streaks from this population simply because these objects are too faint to be seen optically. We also observe a modest bias towards the detection of larger objects, as they are significantly easier to detect than smaller ones by virtue of their size. 

The most distant object observed was an SL-6 R/B, detected at a range of 88\,060\,km, while the closest was the STARLINK‑2366 satellite, observed at a range of 359\,km. A GSLV R/B was detected only 12 days after launch, making it the ‘youngest’ object in our sample, whereas the VANGUARD R/B, launched on 17 March 1958, was observed after nearly 65 years in orbit.

\begin{figure}[h!]
    \centering
    \includegraphics[width=\linewidth]{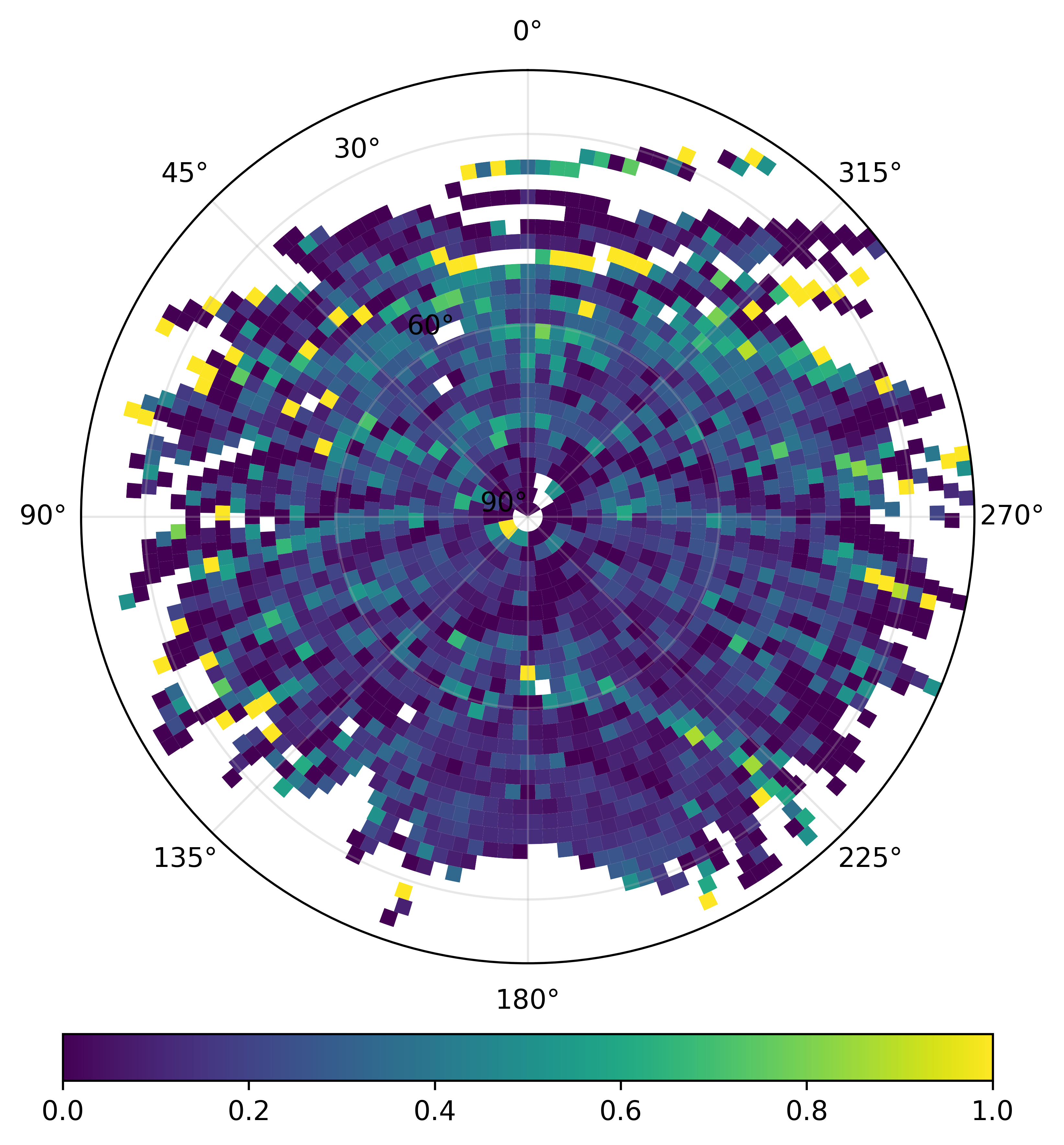}
    \caption{Fraction of images affected by at least one streak for all pointing directions of OmegaCAM images acquired in 2023. The concentric circles represent the altitude in degrees, while the radial axis shows the azimuth in degrees.}
    \label{fig:telescope_pointing}
\end{figure}

\begin{figure}[h!]
    \centering
    \includegraphics[width=\linewidth]{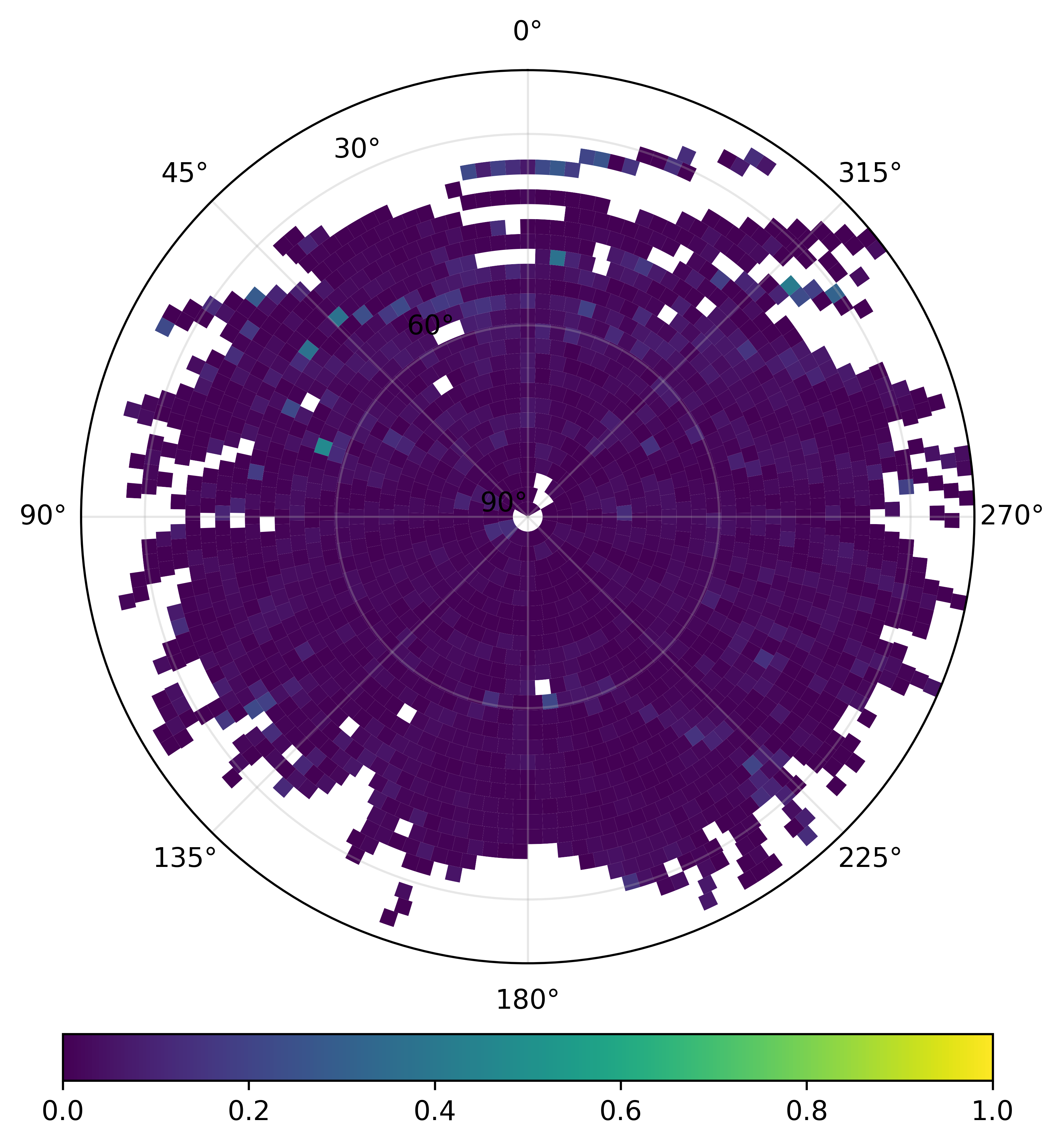}
    \caption{Same as Fig.~\ref{fig:telescope_pointing} for the fraction of affected area. A CCD is counted as affected if at least one streak crosses it, and the fraction is computed as the total affected CCD area divided by the total area of all images in each bin.}
    \label{fig:telescope_pointing_area}
\end{figure}

The pointing directions of all images captured with OmegaCAM in 2023 are shown in Fig.~\ref{fig:telescope_pointing}. The distribution reveals that streak contamination affected nearly every surveyed region of the sky. Over this one-year period, 16.9\% of the images contained at least one streak; however, a closer inspection at the CCD level shows that only 1.9\% of CCDs were affected by a streak, as illustrated in Fig.~\ref{fig:telescope_pointing_area}.

\section{Discussion}\label{sec:discussion}
The results of this study are highly promising. Our detection algorithm has demonstrated the capacity to identify a broad range of streaks, including very faint ones. However, the less favourable results regarding the algorithm's generalisation to real-world data illuminate an important challenge when working with large datasets. In particular, even with a careful selection of training and validation datasets by incorporating diverse backgrounds, moon illumination conditions, seeing variations, and artifacts from images drawn from across the archive and all filters, there is such extensive variation among them that it would be impossible to incorporate all conditions the algorithm will encounter when processing the data. An archive such as OmegaCAM, spanning nearly 15 years of data acquisition, contains an enormous breadth of variety. A single month of archival data would already exceed the size of our selected training set, and the cost of manual annotation would make it infeasible to include more images. Then, despite the careful design of the artificial-streak generator, the diversity of real streak morphologies remains challenging to reproduce in full, while the model inevitably suffers from this mismatch between simulation and reality. 

Several strategies could help mitigate these issues. One approach is to select many more images from the archive to better represent the diversity of backgrounds, while only annotating carefully chosen subregions, since most pixels will not contain streaks. Another is to improve the artificial-streak generator; for example, by incorporating object rotation and more realistic variability in morphology and brightness. A further option is to augment the training set with real streaks detected in an initial run of the algorithm; however, this carries the risk of biasing the model towards features it already detects with a high level of accuracy. Finally, we could combine multiple detection techniques, such as the Hough transform or matched filtering, to perform an extensive initial search for faint or unusual streaks, review and curate these candidates, and then use the resulting high-quality dataset to train the model.

As with many machine-learning methods, our detection algorithm produces a relatively high number of false positives. The observation that most false positives fall into a few well-defined classes (diffraction spikes, filter fringes, saturation bleeds) motivated the development and use of a dedicated classifier. This strategy has proven effective, as the classifier correctly rejects the majority of false detections produced by the detection network. However, as is typical for such problems, we must face a trade-off between removing as many false positives as possible and retaining genuine streaks, whereby some true positives are inevitably lost.

The classifier performs particularly well for the three main false-positive classes, but struggles more with the fourth (miscellaneous) class. This behaviour is expected, since this category groups rarer events and artefacts that are often tied to specific telescope issues or complex image defects, which are difficult to characterise even for human annotators.

One of the most important findings in our study is that roughly 20\% of detections cannot be correlated with objects listed in the catalogue. Many of these uncorrelated streaks appear as faint traces and may correspond to small, uncatalogued debris. This population of small debris is notoriously difficult to track yet represents a real threat to space assets. Cataloguing such objects using large field-of-view telescope archives is therefore a promising direction for future works.

Although the seeing effect produces a visible wobble in OmegaCAM streaks, we do not account for it explicitly during refinement, as fitting a straight line through the Moffat-profile centres effectively averages out these deviations. The refined start- and endpoints therefore correspond to the centroid of a stationary source measured over the portion of the exposure during which the satellite crossed the image. This approximation is justified by the typical observing conditions at ESO Paranal: the 50th percentile seeing FWHM is 0.72 arcsec and values of up to 1.3 arcsec have been reported under poor conditions \citep{sarazin_seeing_2008}. In OmegaCAM pixel units, this corresponds to a typical wobble amplitude of about 3.5 to 6.5 pixels. The atmospheric coherence time $\tau_0$ at ESO Paranal is typically 3.9 ms \citep{sarazin_statistics_2002} and, assuming a centroid jitter timescale that is roughly 20 times longer, gives a streak wobble period of about 78 ms. A satellite on a circular orbit at 400 km altitude crosses the shorter 2000-pixel side of a single OmegaCAM CCD in about 100 ms, so the refinement procedure still improves the astrometric accuracy of the endpoints even in fairly extreme cases. Only edge-grazing streaks from LEO satellites may be affected by imperfect alignment, since in those cases the wobble period can be longer than the observed track segment. The streak alignment procedure is performed on the rotated image with the streak horizontally aligned. This reduces the complexity of locating the streak centres within rectangular apertures and therefore lowers the computational cost. The interpolation and correlated-noise effects introduced by the rotation, which lead to sub-pixel errors in the fitted Moffat centres, correspond to a few tenths of an arcsecond. For a general comparison, a high-quality dedicated optical SSA telescope system is required to provide astrometric accuracy better than 3 arcsec \cite{bloom_space_2022}, which means that even the endpoints of streaks affected by imperfect alignment can still be considered high-quality astrometric measurements.

The analysis of one year of OmegaCAM data provides valuable insight into the performance of the detection pipeline under real observing conditions. It is reassuring that the overall detection statistics align well with the space-track catalogue. The under-detection of small debris in LEO can mostly be attributed to the fact that a telescope such as the VST with its OmegaCAM imager cannot match the sensitivity of radar systems to small, nearby objects. Despite this limitation, the pipeline still detects thousands of streaks, which can then be processed with advanced photometric tools to extract lightcurves, derive rotation rates, measure brightness, and potentially constrain material composition. A forthcoming paper will describe these methods in detail.

As the number of satellites and debris in orbit continues to grow, it becomes increasingly important for astronomers to quantify precisely the impact of streaks on their data. This study shows that no region of the sky is free from satellite-induced optical interference. However, these results alone do not provide a complete picture. A more detailed assessment of the impact on the pixel level is required. Because the algorithm outputs heatmaps of the pixel-wise probability of belonging to a streak, we can (after choosing an appropriate segmentation threshold) either remove streaks from the images or quantify their impact by measuring exactly which pixels they affect. An additional, crucial step will be to study the temporal evolution of the situation by analysing more archival data over many years.

\begin{acknowledgements}
The authors thank Arthur Devaux, Aymeric Deslarzes and Seifeddine Ajmi for their contribution to the annotation of the datasets. We also thank Léonard Flückiger for his initial study on the modelling of atmospheric scintillation. This work has received funding from the Swiss National Science Foundation and the Swiss Innovation Agency (Innosuisse) via the BRIDGE Discovery grant 40B2-0 194729.
\end{acknowledgements}

\bibliographystyle{aa} 
\bibliography{bibliography} 
\begin{appendix}
\onecolumn
\section{Detection algorithm architecture}\label{sec:app_htlcnn_schema}
\begin{figure*}[h!]
    \centering
    \begin{subfigure}[b]{\linewidth}
        \centering
        \includegraphics[width=0.32\linewidth]{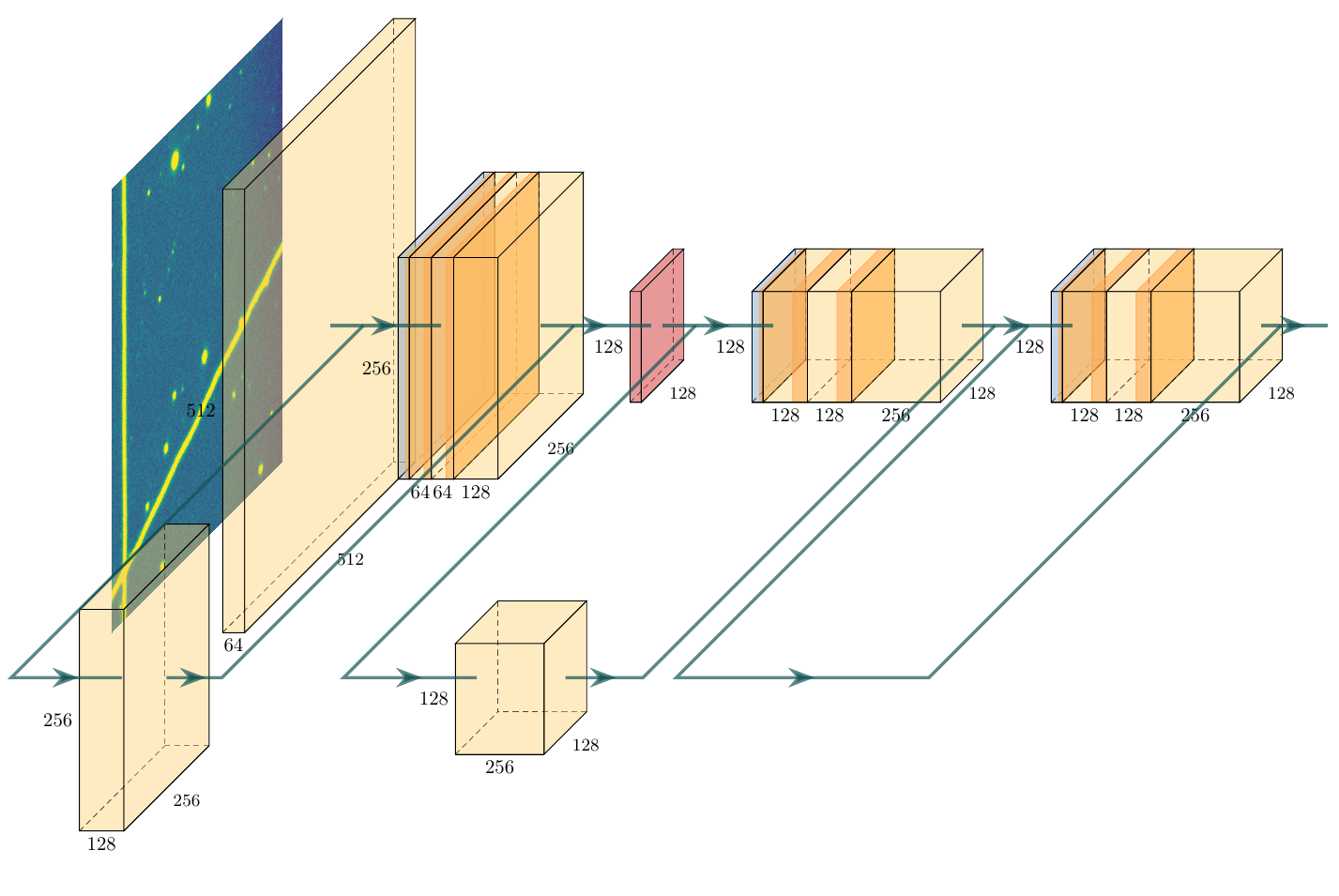}
        \caption{}
        \label{fig:htlcnn_backbone}
    \end{subfigure}
    \begin{subfigure}[b]{\linewidth}
        \centering
        \includegraphics[width=\linewidth]{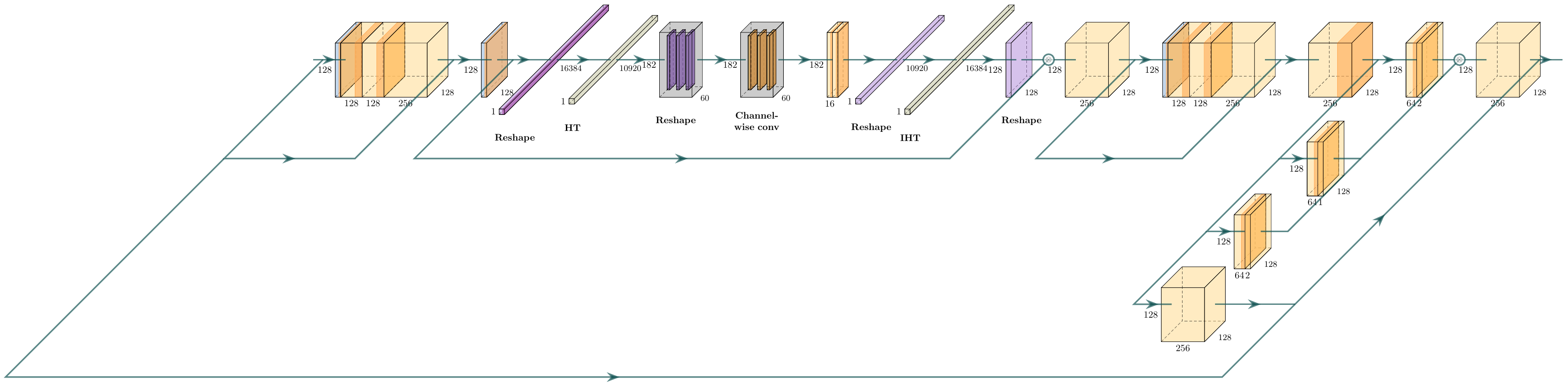}
        \caption{}
        \label{fig:htlcnn_123}
    \end{subfigure}
    \begin{subfigure}[b]{\linewidth}
        \centering
        \includegraphics[width=\linewidth]{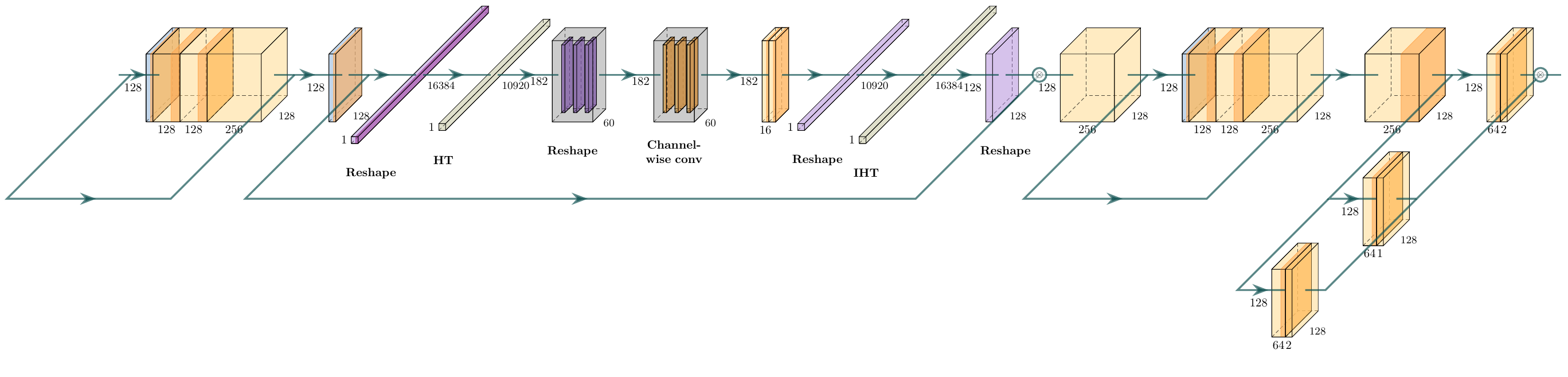}
        \caption{}
        \label{fig:htlcnn_4}
    \end{subfigure}
    \begin{subfigure}[b]{\linewidth}
        \centering
        \includegraphics[width=0.32\linewidth]{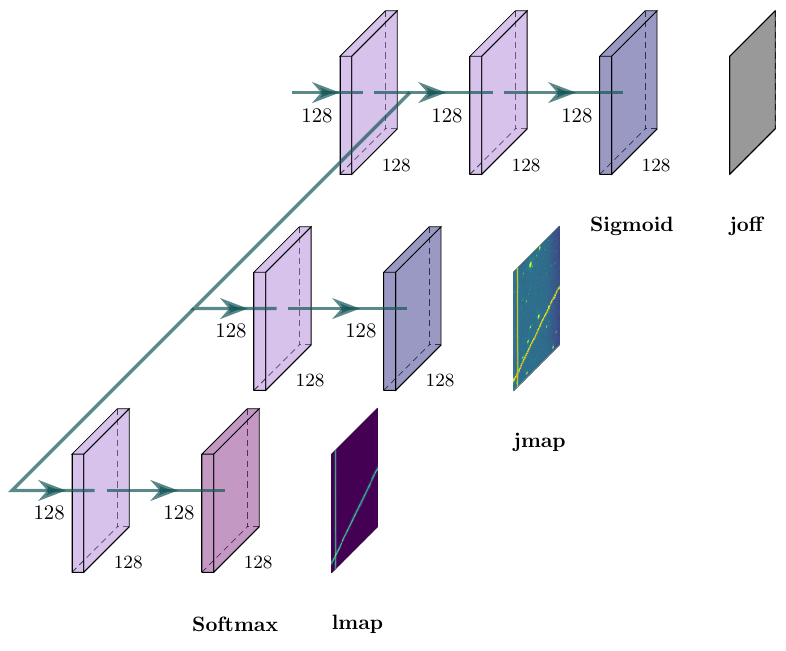}
        \caption{}
        \label{fig:htlcnn_output}
    \end{subfigure}
    \caption{HT-LCNN network architecture. The input image is first processed by module (a), followed by three successive passes through module (b), then a single pass through module (c), and finally through module (d), which produces three separate output heads (jmap, lmap and joff). Yellow boxes denote convolutional layers, with darker box segments indicating ReLU activations. Red layers represent max-pooling operations, blue boxes indicate batch normalization, green boxes denote multiplications with the Hough transform or inverse Hough transform matrices, and purple boxes represent reshaping operations. In module (d), the pink box leading to the lmap output corresponds to a softmax activation function, while the two purple-blue boxes leading to the joff and jmap outputs are sigmoid activations.}
    \label{fig:htlcnn_architecture}
\end{figure*}
\twocolumn

Figure~\ref{fig:htlcnn_architecture} shows the global architecture of the HT-LCNN detection network. In the current implementation, the junction map (jmap) and junction offset (joff) outputs are used together to determine the pixel locations of line endpoints. The final line endpoints are obtained by first detecting junction candidates as local maxima in the junction heatmap jmap, then refining each selected junction to sub-pixel precision using the offset field joff, and finally forming all pairwise junction connections as candidate segments. Each candidate line is uniformly sampled along its length, the extracted features are scored by an MLP-style verification network, and segments that pass the score threshold (> 0.5) are kept as the final predicted lines, with their endpoints rescaled from the 128$\times$128 pixel network grid to the 512$\times$512 pixel image space. The 128$\times$128 pixel line map (lmap) is used only as a training supervision signal, and we solely employ it for visual inspection of the results. For further details on the network implementation and line candidate extraction, refer to the original paper by \cite{lin_deep_2020}.

\section{Derivation of the brightness of space objects for the generation of artificial streaks}\label{sec:app_snr_derivation}
We generate an artificial population of space object by selecting size--altitude pairs from Table~\ref{tab:list_size_alt}.

\begin{table}[h!]
  \begin{center}
    \caption{Object sizes and altitudes used to generate the artificial space object population.}
    \begin{tabular}[h]{cc}
    \hline\hline
    Size [m] & Altitude [km]\\
    \hline
    0.01 & 300 \\  
    0.02 & 350 \\  
    0.03 & 400 \\  
    0.04 & 450 \\  
    0.05 & 500 \\  
    0.06 & 550 \\  
    0.07 & 600 \\  
    0.08 & 650 \\  
    0.09 & 700 \\  
    0.10 & 750 \\  
    0.15 & 800 \\  
    0.20 & 850 \\  
    0.25 & 900 \\  
    0.30 & 950 \\  
    0.40 & 1000 \\  
    0.50 & 1100 \\  
    0.60 & 1200 \\  
    0.70 & 1300 \\  
    0.80 & 1400 \\  
    1.00 & 1500 \\  
    1.50 & 2000 \\  
    2.00 & 3000 \\  
     & 4000 \\  
     & 5000 \\  
     & 10000 \\  
     & 20000 \\  
     & 35786 \\  
     & 36100 \\
    \end{tabular}
    \label{tab:list_size_alt}
  \end{center}
\end{table}

Assuming a spherical shape for all objects (using the sizes from Table~\ref{tab:list_size_alt} as diameter) and a Lambertian scattering of light, we determine the apparent brightness of an object with 
\begin{multline}
    m_{sat} = M_{\odot} - 2.5 \cdot log_{10}(A \cdot \rho \cdot \phi(h)) \\ + 5 \cdot log_{10}(h) + A_\nu \cdot \chi(z).
\end{multline}
$M_{\odot}$ is the Sun's magnitude used as reference, $A$ the cross-sectional area of the object, $\rho$ the object's albedo for which a value of 0.175 is assumed \citep{mulrooney_investigation_2008}, $h$ the altitude of the object, $\phi(h)$ the phase function, computed as
\begin{equation}
    \phi(h) = \frac{1 - \cos(\alpha)}{2},
\end{equation}
with
\begin{equation}
    \cos{(\alpha)} = \frac{R_\oplus + h}{\sqrt{((R_\oplus + h) ^2 + h^2)}},
\end{equation}
and $R_\oplus$ the radius of the Earth. $A_\nu$ is the extinction in the specific filter band\footnote{\url{https://www.eso.org/sci/observing/tools/Extinction.html}} and $\chi(z)$ the optical pathlength along a line of sight in units of air masses, as a function of the zenith angle $z$
\begin{equation}
    \chi(z) = \frac{1}{\sqrt{1 - 0.96 \cdot \sin{(z)} ^2}}.
\end{equation}

From the apparent magnitude $m_{sat}$, we compute the flux of the space object relative to a reference star with a magnitude of 20 ($m_\star$) and its corresponding flux value ($F_\star$) provided by ESO for each filter\footnote{\url{https://www.eso.org/observing/etc/bin/gen/form?INS.NAME=OMEGACAM+INS.MODE=imaging}}
\begin{equation}
    F_{sat} = F_\star \cdot 10 ^{-0.4 \cdot (m_{sat} - m_\star)}.
\end{equation}

We then obtain the total flux contribution of the space object by multiplying this flux by the exposure time of the streak. To account for seeing effects, we calculate the signal within the seeing aperture as
\begin{equation}
    n_{aper} = \frac{n_{sat} \cdot 2 \cdot r_{FWHM}}{l_{streak} + 2 \cdot r_{FWHM}},
\end{equation}
with $n_{sat}$ the total flux contribution from the object, $r_{FWHM}$ the radius corresponding to the FWHM of the PSF and $l_{streak}$ the length of the streak. We finally compute the S/N as
\begin{equation}
    S/N = \frac{n_{aper}}{\sqrt{(n_{aper} + n_{sky} \cdot p + p \cdot \sigma_R^2)}},
\end{equation}
with $n_{sky}$ as the sky background flux, $\sigma_R$ the readout noise of the imager, and $p$ the area factor defined as
\begin{equation}
    p = 2 \cdot r_{FWHM}^2.
\end{equation}

\section{Simulation of atmospheric scintillation}\label{sec:app_scintillation}
The Langevin equation in one dimension is given by
\begin{equation}
    \frac{dx}{dt} = -\frac{\partial V(x)}{\partial x} + \eta(t),
\end{equation}
where $V(x)$ is the potential function and $\eta(t)$ is a stochastic noise term with the following properties, 
\begin{equation}
    \langle \eta(t) \rangle = 0 \quad \text{and} \quad \langle \eta(t) \eta(t') \rangle = 2D\delta(t - t'),
\end{equation}
where $D$ is the diffusion coefficient. In this setting, the stationary probability density of the Langevin dynamics may be represented as
\begin{equation}
    P_{L}(x) = e^{-\frac{V(x)}{D}}.
\end{equation}
As seeing effects can be well represented by a Moffat function, we require the stationary density to follow a Moffat profile,
\begin{equation}
    f(x; \alpha, \beta) = \frac{\beta - 1}{\pi \alpha^2} \left(1 + \frac{x^2}{\alpha^2}\right)^{-\beta},
\end{equation}
where $\alpha$ and $\beta$ are the Moffat scale and shape parameters, respectively. Setting $P_{L}(x) \equiv f(x;\alpha,\beta)$ leads to the potential being expressed (up to an additive constant) as
\begin{equation}
    V(x) = D \ln\left(1 + \frac{x^2}{\alpha^2}\right).
\end{equation}

In two dimensions, this can be generalised to
\begin{equation}
    \frac{d\vec{x}}{dt} = -\nabla V(\vec{x}) + \vec{\eta}(t)
,\end{equation}
where
\begin{equation}
    V(\vec{x}) = D \ln\left(1 + \frac{x^2 + y^2}{\alpha^2}\right)
\end{equation}
and
\begin{equation}
    \nabla V(\vec{x}) = D \cdot \frac{2\vec{x}}{\alpha^2 \left(1 + \frac{x^2 + y^2}{\alpha^2}\right)}.
\end{equation}

By combining these equations with the computed S/N from Appendix~\ref{sec:app_snr_derivation}, we can simulate realistic streaks that include scintillation-induced intensity variations and seeing-induced blurring.

The parameters $\alpha$ (scale parameter) and $\beta$ (shape parameter) of the Moffat distribution are derived from the seeing as
\begin{equation}
    \alpha = \frac{FWHM}{2 \cdot \sqrt{2^{(1 / \beta)} - 1}},
\end{equation}
and we selected a standard value of $\beta=2.5$ for our model \citep{trujillo_effects_2001}.

Since we could not identify a direct correlation between the diffusion coefficient $D$ in the Langevin dynamics model and the physical parameters of atmospheric scintillation, we performed a fitting procedure using multiple simulated images. We generated images with varying diffusion coefficients and measured the wobbling of the streaks in each image. Assuming circular orbits for the objects, we derived their velocities and, using known telescope resolution, aperture, and local wind speed values, calculated their altitudes. We derive the orbital velocity $v$ from the measured wobble $w$ and the scintillation timescale $\tau$ \citep{nir_optimal_2018} using
\begin{equation}
    v = \frac{w}{\tau},
\end{equation}
with
\begin{equation}
    \tau = \frac{D_{t}}{v_{w}},
\end{equation}
where $D_{t}$ is the telescope aperture and $v_{w}$ the wind speed. We compute the object's altitude as
\begin{equation}
    h = \sqrt[3]{\frac{\mu_\oplus}{(v \cdot \theta)^2}}-R_\oplus,
\end{equation}
with $\theta$ as the telescope resolution and $\mu_\oplus$ the standard gravitational parameter of the Earth.

After slight adjustments to the altitude-diffusion coefficient pairs through visual inspection, we derived the following relationship between the altitude $h$ of an object on a circular orbit and its diffusion coefficient, $D$,
\begin{equation}
\label{eq:diffusion_coefficient}
    D = 12 \cdot \log{(h)} - 65.
\end{equation}
We note that h is given in [km] in Eq.~\eqref{eq:diffusion_coefficient}.

\begin{figure}
    \centering
    \includegraphics[width=\linewidth]{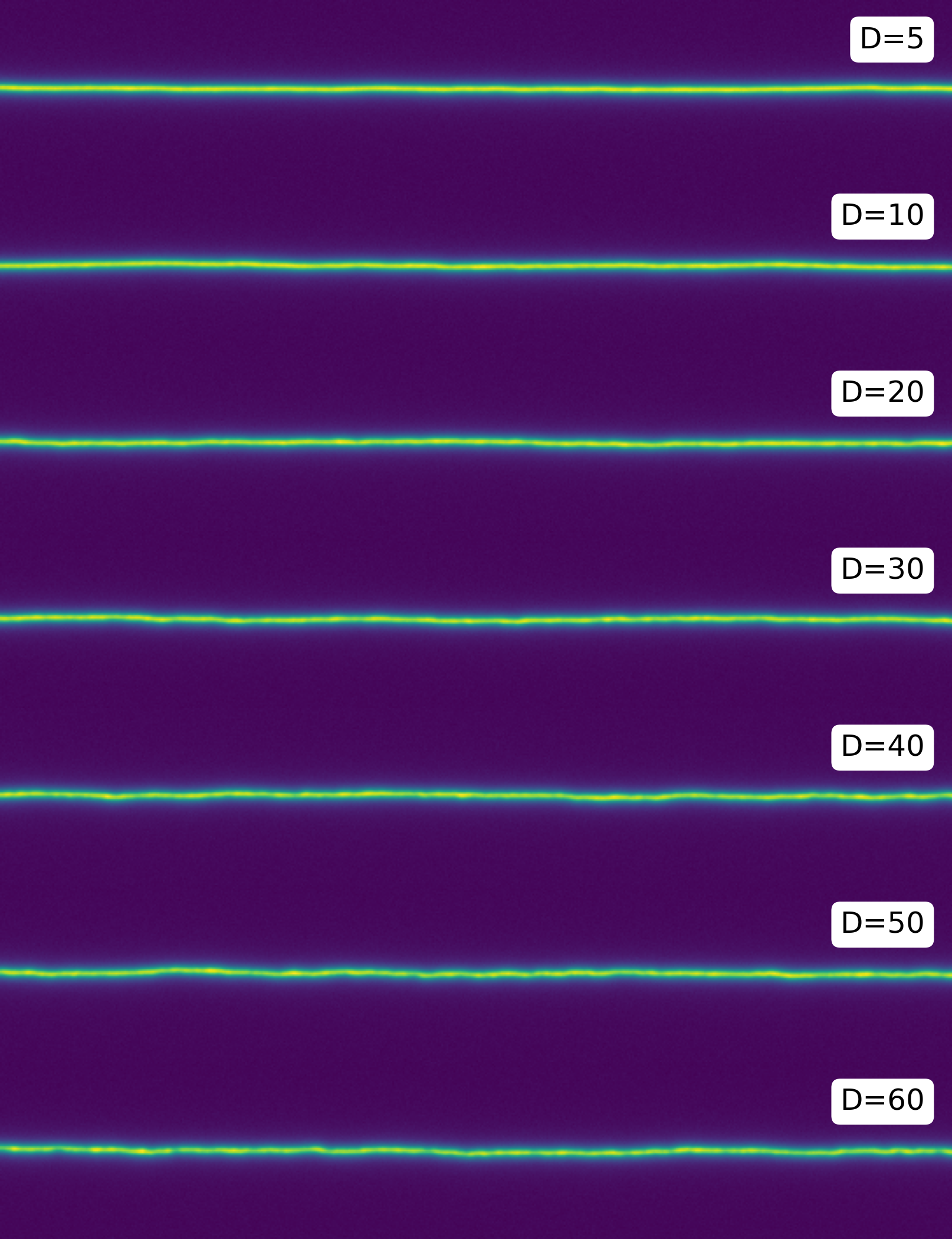}
    \caption{Streaks with varying diffusion coefficient. $D=5$ corresponds to an object in LEO and $D=60$ to an object in GEO.}
    \label{fig:wobbling}
\end{figure}

Sample streaks with variable diffusion coefficients are presented in Fig.~\ref{fig:wobbling}.

\section{Parameter selection for the classification algorithm}\label{sec:app_classifier}
To choose the optimal input format for the classifier, we study how different rescaling and padding strategies affect the mapping of variable-size streak cutouts to a fixed input resolution. The aim is to preserve as much geometric information as possible while providing the classifier with a consistent input size.

We evaluate two target resolutions: a square format ($144 \times 144$ pixel) and a rectangular format ($640 \times 64$ pixel), matching the average streak cutout shape. For each resolution, we compare three strategies. The \texttt{stretch} strategy resizes cutouts directly to the target resolution without preserving the aspect ratio. The \texttt{side-padding} strategy rescales cutouts to match the target height while preserving the aspect ratio and pads horizontally to reach the target width. The \texttt{symmetric-padding} strategy rescales cutouts isotropically to fit within the target resolution and applies symmetric padding on all sides.

\begin{figure}
    \centering
    \includegraphics[width=\linewidth]{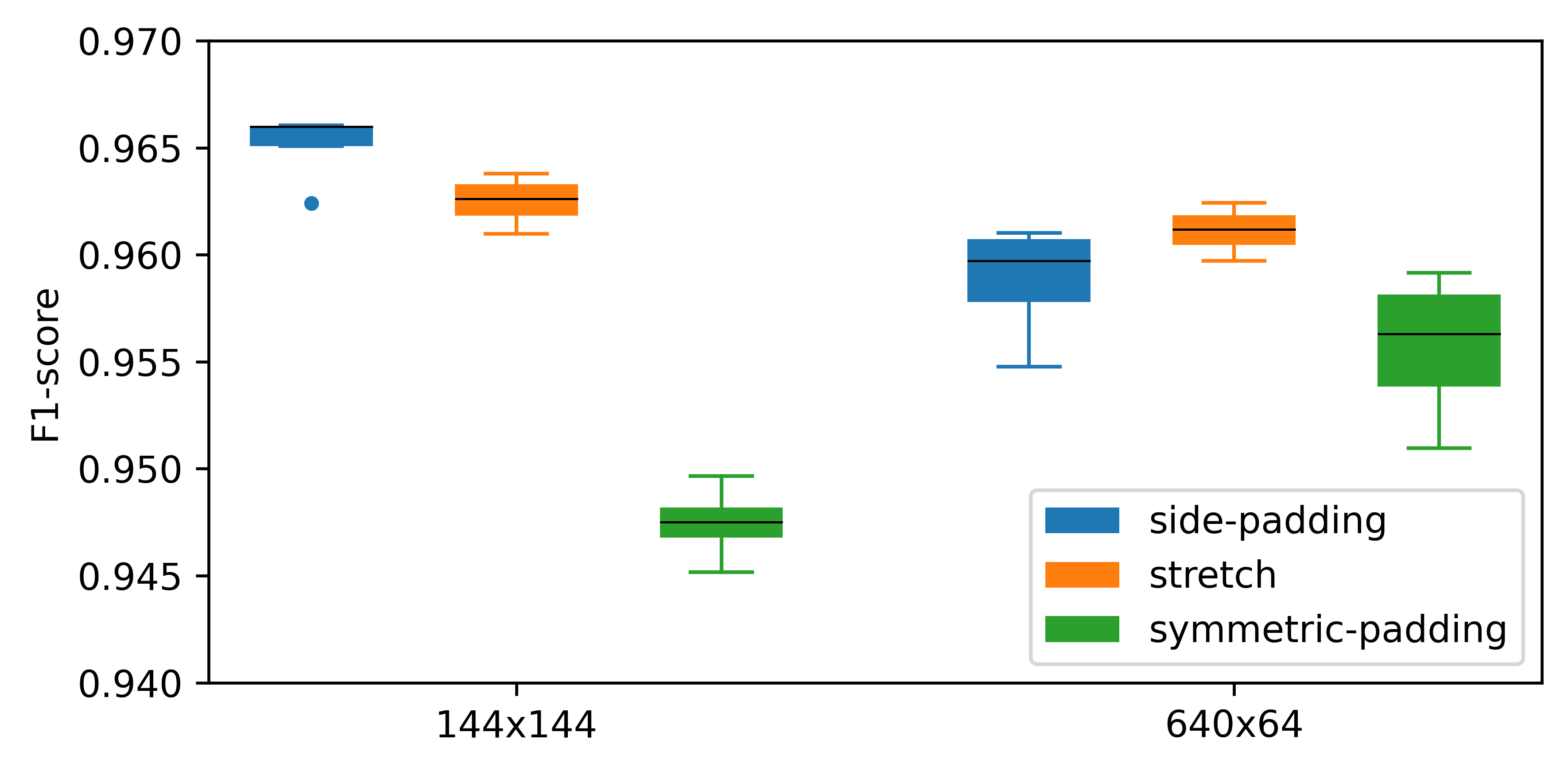}
    \caption{Best validation F1-scores obtained with different resizing and padding strategies for square ($144 \times 144$ pixel) and rectangular ($640 \times 64$ pixel) inputs. Each box summarizes results over three training runs with different random seeds.}
    \label{fig:cutout_format}
\end{figure}

For the $144 \times 144$ pixel resolution, \texttt{side-padding} yields the highest peak validation F1-score (0.965). For the $640 \times 64$ pixel resolution, \texttt{stretch} performs best (0.961), as shown in Fig.~\ref{fig:cutout_format}. Despite the better alignment of the rectangular format with the streak geometry, the square $144 \times 144$ pixel representation with \texttt{side-padding} achieves superior overall performance.

Resizing to a compact square representation may implicitly perform a kind of geometric pooling, concentrating the streak signal while reducing noise to create more stable, discriminative patterns. We therefore adopt this configuration for its higher performance and simpler, more generalisable setup.

\section{Computation time}\label{sec:app_time}
We evaluated the computation time required to run the detection and classification algorithms. Both ran on a desktop computer equipped with an NVIDIA GeForce RTX 3090 Ti GPU (24 GB VRAM).

As shown by the computation times in Table~\ref{tab:time_detection}, inference dominates at 69\% of the total processing time for the detection procedure. This highlights the critical role of GPU acceleration in machine learning applications, especially for processing entire astronomical archives as quickly as possible. The higher relative variability in stitching mostly arises from fluctuating detection counts per mosaic.

\begin{table}[h!]
    \centering
    \caption{Average computation time and standard deviation for the detection algorithm.}
    \begin{tabular}{c|c}
    \hline \hline
        Step & Time [s]\\
    \hline
        Preprocessing & $13.36 \pm 0.78$ \\
        Inference & $32.91 \pm 0.45$\\
        Line stitching & $1.73 \pm 0.97$\\
    \hline
        Total & $48.00 \pm 2.20$
    \end{tabular}
    \tablefoot{Evaluated on one full OmegaCAM mosaic ($16\,000 \times 16\,000$ pixel).}
    \label{tab:time_detection}
\end{table}

For the classification stage, the computational cost per candidate streak is modest, with the total time per cutout dominated by the CPU-bound cutout generation step rather than the CNN inference itself (Table~\ref{tab:time_classification}). Even for datasets with tens of thousands of candidates, the additional wall-clock time incurred by the classifier remains small compared to the initial detection pass.

\begin{table}[h!]
    \centering
    \caption{Average computation time and standard deviation for the classification algorithm.}
    \begin{tabular}{c|c}
    \hline \hline
        Step & Time [s]\\
    \hline
        Cutout generation & $0.24 \pm 0.06$\\
        Inference & $0.03 \pm 0.00$\\
    \hline
        Total & $0.27 \pm 0.06$
    \end{tabular}
    \tablefoot{Evaluated on one streak cutout.}
    \label{tab:time_classification}
\end{table}
\end{appendix}
\end{document}